%% file: clouds_and_bridges.tex
\documentclass[a4,useAMS,usenatbib]{mnras}

\usepackage{graphicx}
\usepackage{setspace}
\usepackage{natbib}
\usepackage{color}
\usepackage{amsmath,amssymb}
\usepackage{times}
\usepackage{aas_macros}
\usepackage{hyperref}

\newcommand{\Gaia}{{\it Gaia}}
\newcommand{\Amp}{{\rm Amp}}
\newcommand{\AEN}{\log_{10}\left({\rm AEN}\right)}
\newcommand{\LMS}{L_{\rm MS}}
\newcommand{\BMS}{B_{\rm MS}}
\newcommand{\LMB}{X_{\rm MB}}
\newcommand{\BMB}{Y_{\rm MB}}

\title[Magellanic Clouds, Streams and Bridges with \Gaia\ DR1]{Clouds,
  Streams and Bridges. Redrawing the blueprint of the Magellanic
  System with \Gaia\ DR1}

\author[Belokurov et al]{Vasily
  Belokurov$^{1}$\thanks{E-mail:vasily@ast.cam.ac.uk}, Denis
  Erkal$^{1}$, Alis J. Deason$^{2}$ Sergey E. Koposov$^{1}$,
  \newauthor Francesca De Angeli$^{1}$, Dafydd Wyn Evans$^{1}$,
  Filippo Fraternali$^{3}$, Dougal Mackey$^{4}$\\$^{1}$Institute of
  Astronomy, Madingley Rd, Cambridge, CB3 0HA\\$^{2}$Institute for
  Computational Cosmology, Department of Physics, University of
  Durham, South Road, Durham DH1 3LE, UK\\$^{3}$Department of Physics
  and Astronomy, University of Bologna, viale Berti Pichat 6/2,
  I-40127 Bologna, Italy\\$^{4}$Research School of Astronomy and
  Astrophysics, Australian National University, Canberra, ACT 2611,
  Australia}

\begin{document}


\maketitle

\label{firstpage}

\begin{abstract}
We present the discovery of stellar tidal tails around the Large and
the Small Magellanic Clouds in the \Gaia\ DR1 data. In between the
Clouds, their tidal arms are stretched towards each other to form an
almost continuous stellar bridge. Our analysis relies on the exquisite
quality of the \Gaia's photometric catalogue to build detailed
star-count maps of the Clouds. We demonstrate that the \Gaia\ DR1 data
can be used to detect variable stars across the whole sky, and in
particular, RR Lyrae stars in and around the LMC and the
SMC. Additionally, we use a combination of \Gaia\ and {\it Galex} to
follow the distribution of Young Main Sequence stars in the Magellanic
System. Viewed by \Gaia, the Clouds show unmistakable signs of
interaction. Around the LMC, clumps of RR Lyrae are observable as far
as $\sim20^{\circ}$, in agreement with the most recent map of
Mira-like stars reported in \citet{alis_mira}. The SMC's outer stellar
density contours show a characteristic S-shape, symptomatic of the
on-set of tidal stripping. Beyond several degrees from the center of
the dwarf, the \Gaia\ RR Lyrae stars trace the Cloud's trailing arm,
extending towards the LMC. This stellar tidal tail mapped with RR
Lyrae is not aligned with the gaseous Magellanic Bridge, and is
shifted by some $\sim5^{\circ}$ from the Young Main Sequence
bridge. We use the offset between the bridges to put constraints on
the density of the hot gaseous corona of the Milky Way.

\end{abstract}


\section{Introduction}

``The Magellanic Clouds are a pair of massive dwarf galaxies orbiting
the Milky Way.''  While seemingly obvious on the surface, the
statement above conceals a host of failed observational attempts to
verify its parts. Presently, the jury is still debating whether or not
the Clouds have been bound to the Galaxy for very long, if at all
\citep[][]{Besla2007,nitya2013}. The time they have spent as a binary
is unknown \citep[see e.g.][]{Besla2012,Diaz2012}, and their masses
remain unconstrained, although a coherent picture is starting to
emerge in which the Clouds appear much heavier then previously thought
\citep[see e.g.][]{vdm2014,jorge_lmc}. So far, the only fact
established with certainty is that the LMC and the SMC should not
really be here today \citep[][]{busha_lmc,tollerud_lmc}. Yet luckier
still, the two dwarfs are perfectly positioned for study: close enough
so that their individual stars can be resolved and their proper
motions measured, but not so close that they cover half of the
sky. Having the full view, not just a close-up, is crucial, as the
picture of the Magellanic system is complex and filled with scattered
intricate details that only make sense in concert. Many of these are
in fact signs of the ongoing interaction, both between the Clouds
themselves, and of the pair with the Milky Way.

The first observational example of a morphological feature in the SMC
most likely caused by its larger neighbour is the eastern stellar
``Wing'' discovered by \citet{Shapley1940}. The second clue to the
Clouds' turbulent relationship is the neutral hydrogen ``bridge''
connecting the dwarfs, revealed by the study of
\citet{Hindman1963}. As \citet{Irwin1985} showed, this gaseous
Magellanic Bridge (MB) is a site of recent star-formation, hosting a
river of O and B stars, of which the eastern Wing is just the most
visible portion. According to the subsequent study of
\citet{Irwin1990}, this river continues the better part of the
distance from the Small to the Large Cloud \citep[also
  see][]{Battinelli1992} and possibly even contains a faint
``envelope'' of Population II stars. The existence of the young
stellar bridge connecting the LMC and the SMC has recently been
confirmed by \citet{skowron_bridge} using the OGLE's wide and
continuous coverage of the area. However, the presence of the recently
formed stars in the MB carries little information - besides the
important constraint on the timescale - as to the exact course of the
interaction that pulled a great quantity of HI from the SMC to form
the bridge itself.

All models agree that the most straightforward way to produce the MB
is via tidal stripping of the SMC's gas by the LMC \citep[see
  e.g.][]{Besla2012,Diaz2012,Hammer2015}. However, the ferocity of the
interaction can be dialed up and down, in accordance with the (poorly
known) size and the shape of the SMC's orbit around the LMC. The SMC's
orbit is not tightly constrained because the Cloud's relative proper
motion has remained uncertain, as have the masses of both
dwarfs. Naturally, in the case of a close fly-by, the LMC's tides
would also remove some of the SMC's stars. Therefore, the mere
existence of the stellar tidal tail corresponding to the gaseous MB
may serve as a powerful confirmation of the recent direct
interaction. Furthermore, as successfully demonstrated with other
Galactic satellites \citep[see e.g.][]{Dehnen2004,Fellhauer2007,
  Gibbons2014}, tidal streams can also be used to reveal a lot more
about the orbital evolution of the Clouds.

The two seminal papers describing the interaction of the Clouds,
i.e. \citet{Diaz2012} and \citet{Besla2012}, both predict stellar
tidal tails around the SMC, albeit based on different orbital
solutions for the pair. The simulations of \citet{Diaz2012} follow a
light LMC+SMC pair on multiple passages around the Milky Way. Here,
the dwarfs come together as a pair $\sim 2$ Gyrs ago and therefore
only have enough time for two rendezvous. The most recent encounter
between the Clouds, which in this setup happens some 250 Myr ago,
creates two prominent - both gaseous and stellar - tidal tails on
either side of the SMC. One of these is seen in HI today as the MB and
connects the Clouds on the sky and along the line of sight, while the
other, dubbed by \citet{Diaz2012} the ``counter-bridge'', wraps around
the SMC and stretches to much larger distance, i.e. $\sim 90$ kpc
\citep[see also][]{Muller2007}.

\citet{Besla2012} present the results of two Magellanic
simulations. In both, the MCs are much heavier compared to the model
of \citet{Diaz2012} and have just passed their first pericenter around
the Galaxy. However, contrary to the model of \citet{Diaz2012}, the
Clouds are allowed to interact with each other for a much longer
period. The intensity of this interaction is different for the two
models of \citet{Besla2012}: Model 1 has a large pericentric distance
of the SMC around the LMC, of order of $\sim 30$ kpc, while in Model
2, there is a direct collision between the dwarfs. Accordingly, the
gaseous MB appears rather weak in Model 1 and very dramatic in Model
2. The difference in the MB gas contents in the two models is also
reflected in the distinct star-forming properties of the Bridge: in
Model 1 the density of neutral hydrogen is too low to kick-start star
production, while in Model 2, there is copious in-situ MB star
formation. Note, however, that the amount of the stripped MB stellar
debris in Model 2 does not necessarily match the high gas
density. This is because, during the collision, the SMC's gas is
stripped not only by the LMC's tidal force but also by the ram
pressure of its gaseous disc. Nevertheless, as the follow-up treatise
by \citet{Besla2013} demonstrates, Model 2 predicts a factor of
$\sim$5 more old stellar (e.g. RR Lyrae stars) tidal debris in the MB
compared to Model 1 (see their Table 5).

\begin{figure}
  \centering
  \includegraphics[width=0.48\textwidth]{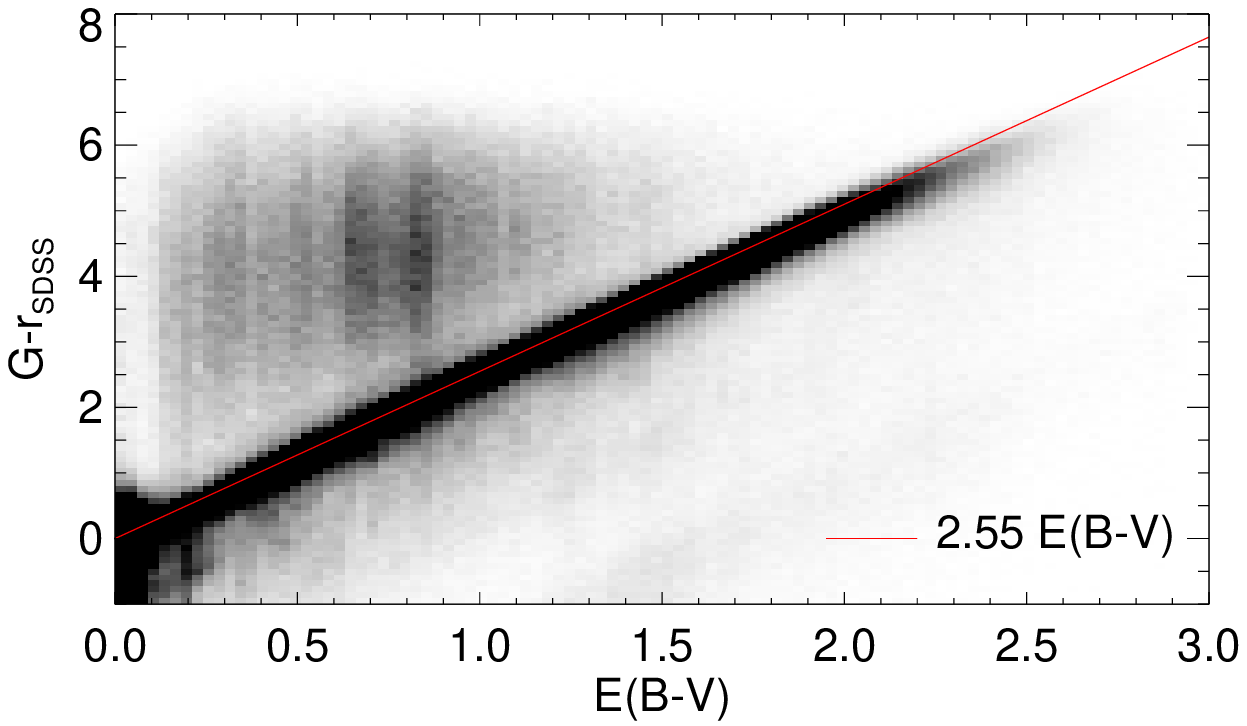}
  \includegraphics[width=0.48\textwidth]{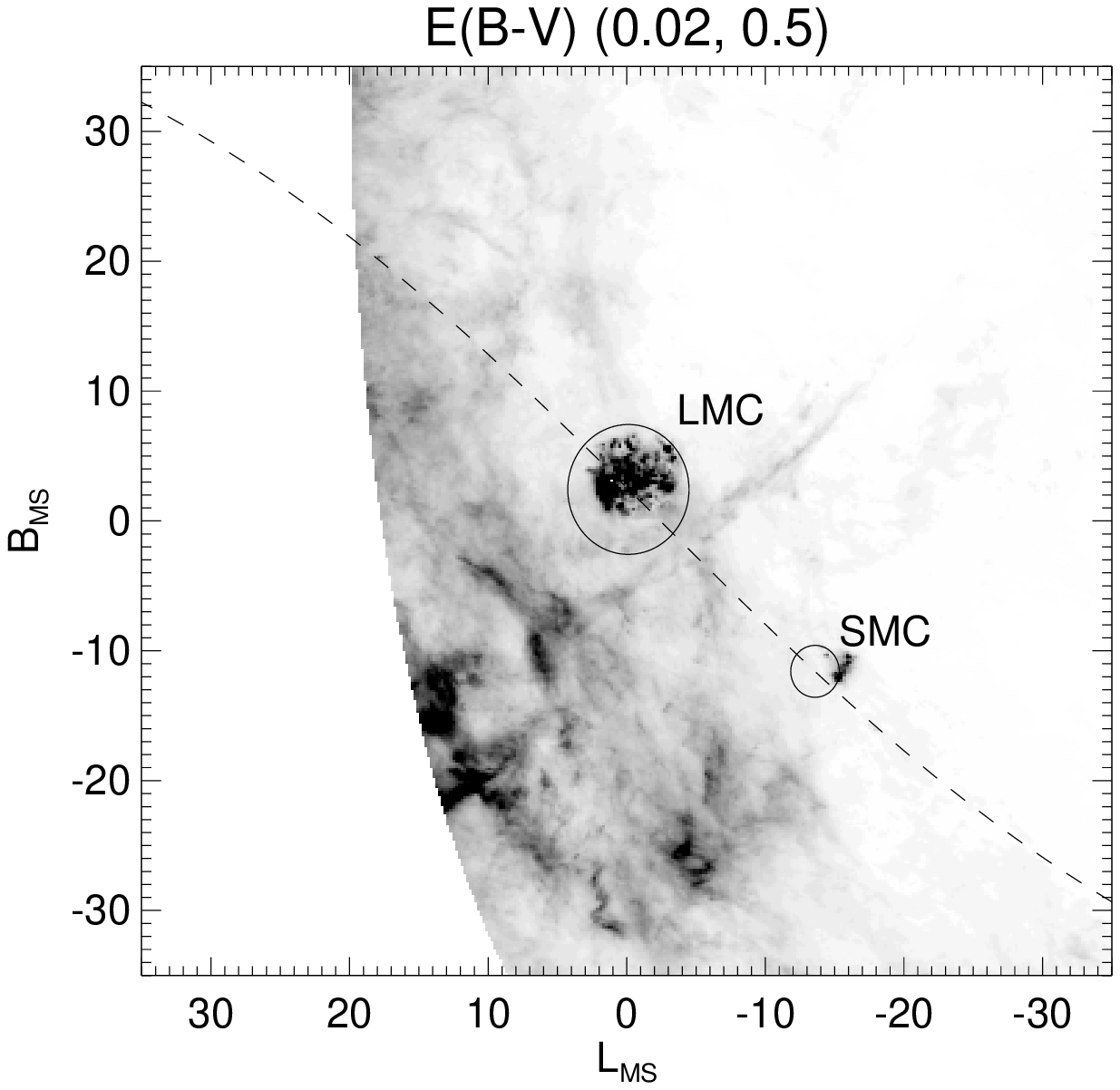}
  \caption[]{\small {\it Top:} Density of stars with both SDSS and
    \Gaia\ measurements and $(g-r)_{SDSS}<0.3$, in the plane of
    $G-r_{SDSS}$ and $E(B-V)$. $G$ is not corrected for dust
    reddening, but $r_{SDSS}$ is. The solid red line shows that
    $A_G=2.55E(B-V)$ provides a reasonable fit to the data. {\it
      Bottom:} The distribution of the dust extinction from
    \citet{SFD} in the Magellanic Stream coordinates for locations
    with Galactic $b<-15^{\circ}$. The black dashed line is the
    equator of the Magellanic Bridge (MB) coordinate system. The black
    circle with a radius of $5^{\circ}$ ($2^{\circ}$) marks the
    location of the LMC (SMC).}
   \label{fig:ebv}
\end{figure}

As both \citet{Diaz2012} and \citet{Besla2012}, as well as a string
of authors before them, predict a stellar counter-part to the gaseous
MB, the region between the Clouds corresponding to the highest HI
column density has been trawled thoroughly for the tidally stripped
SMC stars. The results of the search are somewhat inconclusive. For
example, \citet{Demers1998} and \citet{Harris2007} conclude that the
stellar population of the MB is predominantly young and very little, if
any, stellar material has been stripped from the SMC. On the other
hand, \citet{Kunkel2000, Nidever2013, Bagheri2013,skowron_bridge} and
\citet{Noel2015} all find evidence of an intermediate-age stellar
population in the MB area.

In this Paper, we take advantage of the unique photometric dataset,
provided as part of the \Gaia\ Data Release 1 (GDR1), to study the
low-surface brightness stellar density field around the Magellanic
Clouds. \Gaia\ is the European Space Agency's project to create a
detailed map of the Galactic stellar distribution. \Gaia\ scans the
entire sky constantly, thus providing a record of stellar positions
and fluxes, as well as tangential motions and flux variations over a
period of 5 years with a typical temporal sampling window of $\sim$1
month. \Gaia's limiting magnitude in a very wide optical $G$ band is
$\sim20.5$ which is similar to the limiting magnitude ($I\sim 21$) of
the OGLE IV survey\footnote{\url{http://ogle.astrouw.edu.pl/}}, which
before 14 Sep 2016 provided the widest coverage of the Magellanic
system at this depth.

This Paper is organised as follows. Section~\ref{sec:view} describes
the behaviour of the Galactic dust reddening in the \Gaia\ $G$-band
around the Magellanic Clouds and presents the star-count maps of both
dwarfs. Section~\ref{sec:rrl} introduces the \Gaia\ variability
statistics and discusses how genuine variable stars can be
distinguished from artifacts. We kindly warn the reader that parts of
this Section are rather technical, and those mostly interested in the
properties of the Magellanic Clouds rather than the particularities of
the \Gaia's photometry, might like to skip straight to the next
Section. Section~\ref{sec:bridges} presents the discovery of the
trailing tidal tail of the SMC and a new map of the Young stellar
bridge traced with a combination of \Gaia\ and {\it Galex}. Finally,
Section~\ref{sec:disc} puts the discovery into context.

\section{Magellanic Clouds in \Gaia\ DR1}
\label{sec:view}

The analysis reported below relies on the all-sky source table {\texttt
  GaiaSource} released as part of the \Gaia\ DR1 \citep[see][for
  details]{gaiadr1,gaiafloor}.

\subsection{Extinction}

Before any examination of the \Gaia\ photometry can be carried out,
the apparent magnitudes must be corrected for the effects of Galactic
dust extinction. The LMC's Galactic latitude is only $b \sim
-30^{\circ}$ and there are plenty of filamentary dust patches with
high reddening levels in its vicinity. \Gaia's $G$ band is very wide
and therefore the conversion from $E(B-V)$ to extinction $A_G$ is a
complex function of the source's spectral energy distribution. For the
analysis presented here, this prescription can be simplified as we are
concerned with stars in a narrow range of color, i.e. $0.2<B-V<0.4$
\citep[see e.g.][for details]{Catelan2009}. Based on the pre-flight
simulations, \citet{Jordi2010} recommends $A_G/E(B-V) \sim 3$ for
stars with colors consistent with those of RR Lyrae (see top left
panel of their Figure 17).

\begin{figure}
  \centering
  \includegraphics[width=0.48\textwidth]{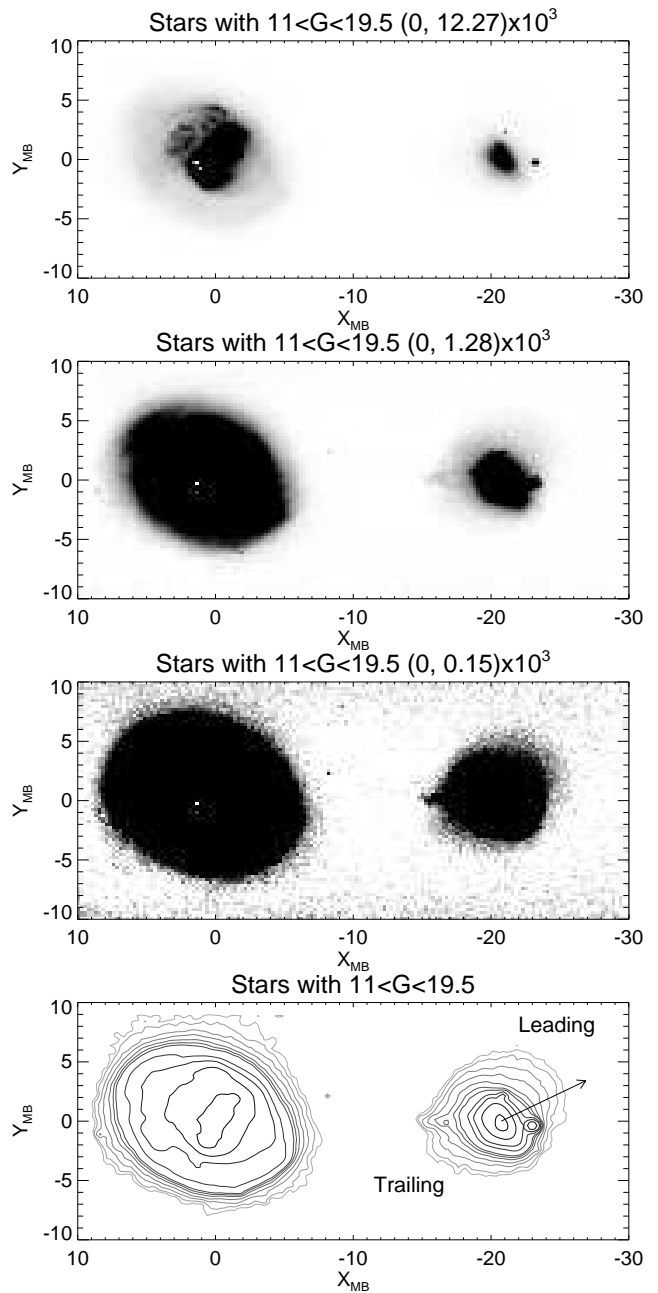}
  \caption[]{\small Density of stars with $11<G<19.5$ in the MB
    coordinate system. The pixels are
    $0.25^{\circ}\times0.25^{\circ}$. A background model comprised of
    a quadratic polynomial for each column (constrained over the range
    of $-13^{\circ}<\LMB<35^{\circ}$, but excluding
    $-8^{\circ}<\LMB<8^{\circ}$) is subtracted. Across all panels, the
    density distribution is the same, only the dynamic range of the
    pixel values changes. The number of stars corresponding to white
    (low counts) and black (high counts) levels is shown in brackets
    in the title of each panel. {\it First (Top) Panel:} Note the
    perturbed disc of the LMC. {\it Second Panel:} Note the low
    surface-brightness extension of the SMC, the 47 Tuc cluster in the
    western part of the dwarf and the Wing in the eastern side, facing
    the LMC. {\it Third Panel:} Note the characteristic S-shape of the
    outer envelope of the density distribution of the SMC. {\it Fourth
      (Bottom) Panel:} Note the twist of the iso-density contours from
    the center outwards. The two protruding ends of the S-shape are
    the origins of the SMC tidal tails.  The arrow shows the Cloud's
    proper motion relative to the LMC as computed using the values
    from \citet{nitya2013}. The Solar reflex motion is subtracted. The
    leading and trailing tails are marked taking into account the
    direction of the SMC's motion around the LMC. }
   \label{fig:starcounts}
\end{figure}

With the GDR1 in hand, it is now possible to estimate the extinction
coefficient, $A_G$, directly from the data. Here, we do it by simply
comparing the uncorrected $G$ magnitude to de-reddened SDSS $r$ band
magnitude for all stars measured by both \Gaia\ and SDSS as a function
of $E(B-V)$. Here, we use the SDSS $r$ band as it is closely related to the \Gaia\ $G$-band.
 The result of this comparison is shown in
Figure~\ref{fig:ebv}, where the following relationship appears to hold
true:

\begin{equation}
  \label{eq:ag}
A_G=2.55E(B-V)  
\end{equation}

\noindent Reassuringly, this is not too far from the value suggested
by \citet{Jordi2010} based on the theoretical calculations. In the
analysis that follows, the $G$-band magnitudes are de-reddened using
the conversion above. We have also checked the behaviour of $A_G$ for
stars in other color regimes, and Equation~\ref{eq:ag} appears to work
well, albeit with increased scatter.

\begin{figure*}
  \centering
  \includegraphics[width=0.98\textwidth]{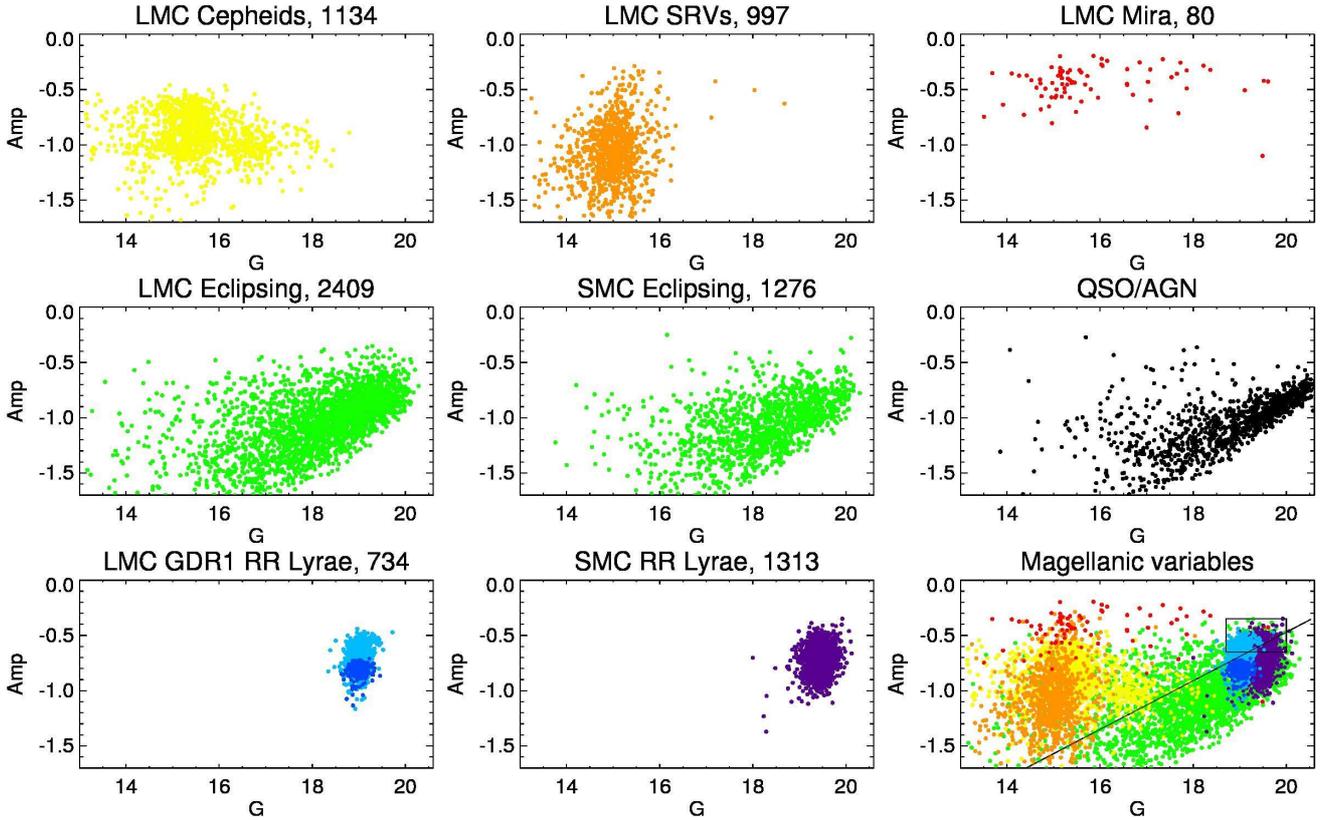}
  \caption[]{\small Variable sources in Gaia DR1. Previously
    identified objects with secure classification are shown in the
    plane spanned by the variability amplitude \Amp\ and the Gaia
    magnitude $G$. The title also reports the number of variables
    shown. {\it Top Left:} LMC Cepheids with
    $3^{\circ}<D_{LMC}<10^{\circ}$ from \citet{cepheid_lmc}. {\it Top
      Center and Right:} Long-period semi-regular variables and Mira
    stars in the LMC from \citet{lpv_lmc} with
    $3^{\circ}<D_{LMC}<10^{\circ}$. {\it Middle Left:} LMC eclipsing
    binaries with $3^{\circ}<D_{LMC}<10^{\circ}$ from
    \citet{eclipsing_lmc}. {\it Middle Center:} SMC eclipsing binaries
    with $D_{SMC}>1^{\circ}$ from \citet{eclipsing_smc}. {\it Middle
      Right:} QSO and AGN from \citet{qso_agn}. {\it Bottom Left:}
    Gaia DR1 RR Lyrae with $5^{\circ}<D_{LMC}<10^{\circ}$
    \citep[see][for details]{rrl_lmc}. RRab (RRc) stars are shown in
    light (dark) blue. {\it Bottom Center:} RR Lyrae in the SMC with
    $D_{SMC}>1^{\circ}$ from \citet{rrl_smc}. {\it Bottom Right:} All
    Magellanic variables from other panels. Black lines show the RR
    Lyrae selection boundaries used in the analysis. This selection
    yields $38\%$ completeness for the LMC RR Lyrae and $13\%$
    completeness for the SMC RR Lyrae. Note that the variables shown
    in this Figure also had to pass the additional cuts detailed in
    Section~\ref{sec:selection}}
   \label{fig:variables}
\end{figure*}

\subsection{Magellanic Stream and Magellanic Bridge coordinate systems}

The bottom panel of Figure~\ref{fig:ebv} displays the Galactic dust
map as calculated by \citet{SFD} in the Magellanic Stream (MS)
coordinates. This coordinate system is suggested by
\citet{nidever2008} and is approximately aligned with the extended
trailing tail of neutral hydrogen emanating from the Clouds. The
Galactic coordinates can be converted into the MS longitude $\LMS$ and
latitude $\BMS$ by aligning with a great circle with a pole at $(l,
b)=(188.5^{\circ},-7.5^{\circ})$. Note that in the MS system, the LMC
lies slightly off-center and has coordinates $(\LMS,
\BMS)=(-0.14^{\circ}, 2.43^{\circ})$, while the SMC is positioned at
$(\LMS, \BMS)=(-15.53^{\circ}, -11.58^{\circ})$.

The dashed line in the bottom panel of Figure~\ref{fig:ebv} indicates
the equator of a different coordinate system in which both the LMC and
the SMC have zero latitude. We call this system the Magellanic Bridge
coordinates as its equator aligns well the HI bridge (see
Section~\ref{sec:hi}). The Equatorial coordinates can be converted
into the MB longitude $\LMB$ and latitude $\BMB$ by aligning with a
great circle with a pole at $(\alpha, \delta)=(39.5^{\circ},
15.475^{\circ})$. In this new Magellanic Bridge coordinate system, the
LMC is at $(\LMB, \BMB)=(0^{\circ}, 0^{\circ})$ and the SMC is at
$(\LMB, \BMB)=(-20.75^{\circ}, 0^{\circ})$.

\subsection{Star-count maps}
\label{sec:starcounts}

\begin{figure*}
  \centering
  \includegraphics[width=0.98\textwidth]{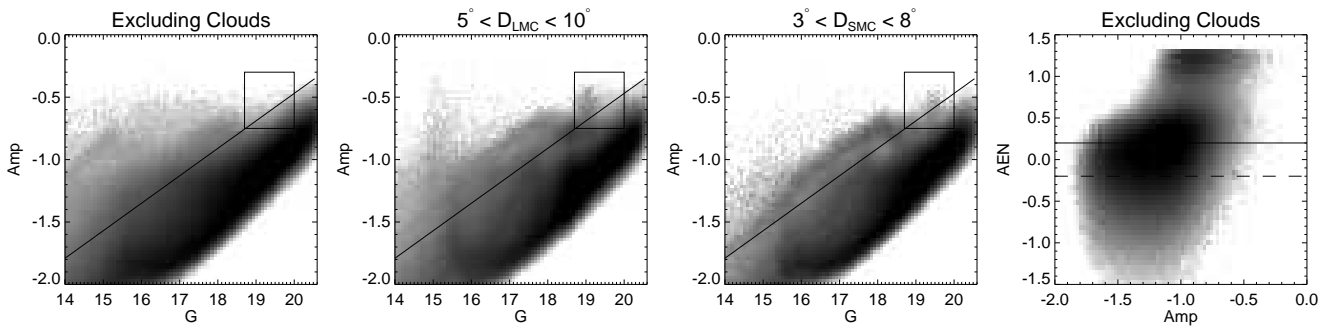}
  \caption[]{\small True and spurious variable objects in Gaia
    DR1. The first three panels show the logarithm of density of
    objects in $(\Amp, G)$ space. The first panel shows stars in
    a $80^{\circ}\times80^{\circ}$ region centered on the LMC, but with
    the area around the LMC and the SMC and below Galactic
    $b=20^{\circ}$ excluded. The second (third) panel presents the stars
    around the LMC (SMC). Stars below the black diagonal line are
    mostly constant, while those above it appear variable
    (approximately 3.9 million in this area) in Gaia DR1. While some
    of the genuine variable stars do cluster in this space, e.g. RR
    Lyrae (see Figure \ref{fig:variables}), mostly, these do not
    produce well-defined sequences spanning large ranges of
    magnitude. Much of the clustering visible in these panels is due
    to spurious ``variables'' likely caused by cross-match
    failures. Note that our final RR Lyrae selection box avoids the
    vast majority of the prominent artifacts visible in the first
    three panels. The fourth panel gives the logarithm of density of stars
    in the space of (the logarithm of) excess astrometric noise, ${\rm
      AEN}$, and variability amplitude, \Amp, for objects with
    $18.7 < G < 20$, i.e. those in the RR Lyrae selection range. The
    stand-alone cloud with large $\log_{10}({\rm AEN})$ is mostly
    galaxies. Our working hypothesis is that the objects with
    cross-match problems appear as spurious photometric variables with
    large ${\rm AEN}$. Therefore we exclude a chunk of $({\rm AEN,
      \Amp})$ space in which the two correlate. Using a conservative
    cut of $\log_{10}({\rm AEN})<0.25$ (black horizontal line) leaves
    2.9 million variable objects, while a strict cut of
    $\log_{10}({\rm AEN})<-0.2$ leaves 1.6 million variable stars.}
   \label{fig:selection}
\end{figure*}

Figure~\ref{fig:starcounts} gives the density distribution of all
stars in the GDR1's {\texttt GaiaSource} table with $11<G<19.5$ in the
MB coordinate system. From top to bottom, the density distribution
remains the same, but the dynamic range of the greyscale image is
varied. The top panel highlights the high density regions, while the
second from bottom panel emphasises the low-surface brightness
environs of the Clouds. Finally, the bottom panel attempts to
summarise the behaviour of the stellar density across all
surface-brightness levels. Note that a simple model of the Galactic
foreground/background has been subtracted from these stellar density
maps. Namely, the number counts in the range
$-13^{\circ}<\LMB<35^{\circ}$, but excluding
$-8^{\circ}<\LMB<8^{\circ}$ were modeled as a quadratic polynomial of
$\BMB$. The parameters of the model were constrained independently for
each pixel column.

As illustrated in the Figure, the LMC star-count distribution harbors
a dense central core with a number of clumpy (presumably star-forming)
regions, surrounded by an irregularly shaped spiral or ring-like
pattern.  The shape and the position angle of the LMC's iso-density
contours evolve from the center outwards. In this map, the LMC can be
seen as far as $\LMB\sim9^{\circ}$ in the East and $\LMB=-8^{\circ}$
in the west. Overall, this view of the LMC appears remarkably similar
to that published recently by \citet{Besla2016}. The first two panels
show several small-scale features in the SMC: most notably, the 47 Tuc
globular cluster in the West and the Wing in the East. The pointy tip
of the Wing at $(\LMB,\BMB)=(-15^{\circ}, 0^{\circ})$ remains a
dramatic feature in all panels of the Figure.

A substantial twist in the SMC's iso-density contours can be seen in
the third (or second from bottom) panel. Here, the outer stellar
density distribution appears to have the characteristic S-shape,
typical of the tidal tails around disrupting satellites \citep[see
  e.g.][]{pal5tails}. In this picture, the SMC tails appear rather
stubby and drop out of sight around $\sim6^{\circ}$ away from the
satellite. The orientation of the two tails seem to be well-aligned
with the SMC's relative (to the LMC) proper motion vector, shown as an
arrow in the bottom panel of the Figure. Given the direction of the
Cloud's motion, we designate the tail stretching towards the top right
or $(\LMB,\BMB)=(-24^{\circ}, 5^{\circ})$ as leading, and the tail
pointing toward the LMC, more precisely towards
$(\LMB,\BMB)=(-16^{\circ}, -3^{\circ})$, as trailing. Note that the
twisting/elongation of the SMC density contours is in broad agreement
with the earlier studies of the dwarf using tracers like Cepheids, Red
Clump stars and RR Lyrae and is intimately linked to its changing
extent along the line of sight \citep[see
  e.g.][]{Scowcroft2016,Nidever2013,ogle_rrl2}.

\section{Magellanic RR Lyrae in \Gaia\ DR1}
\label{sec:rrl}

\subsection{Variable stars in \Gaia\ DR1}
\label{sec:variables}

\Gaia\ is a variable star machine. By scanning the entire sky multiple
times over a baseline of many years, it reveals objects that change
brightness across a wide range of timescales and amplitudes. As
displayed by the sample of the LMC RR Lyrae and Cepheids published as
part of GDR1, the quality of the \Gaia\ lightcurves is exquisite and
is bested only by {\it Kepler}. While the deluge of the all-sky
variability data is expected to be unleashed in the coming data
releases, the GDR1's {\texttt GaiaSource} (GSDR1) table contains
enough information to identify objects whose flux changes with time
and even group them broadly into classes.

Figure~\ref{fig:variables} shows the distribution of previously
identified variables sorted according to their type in the plane of
the variability amplitude statistic \Amp\ (see below) as a function of the
\Gaia\ $G$ magnitude. This variability amplitude estimate relies on the
fact that GSDR1 reports the mean flux as well as the error of the mean
flux estimate. For variable sources, the mean flux error gauges the
range of oscillation in the object's flux. Therefore, for each source
in GSDR1, we can define \Amp\ as follows:

\begin{equation}
  {\rm Amp}=\log_{10}\left(\sqrt{N_{\rm obs}}\frac{\sigma_{\overline{I_G}}}{\overline{I_G}}\right)
\end{equation}

\noindent Here, $N_{\rm obs}$ is the number of CCD crossings,
$\sigma_{\overline{I_G}}$ is the mean $G$ flux error and
$\overline{I_G}$ is the mean $G$-band flux. Figure~\ref{fig:variables}
presents the GSDR1 view of several of the familiar classes of variable
stars residing in the Magellanic Clouds, such as the LMC Cepheids
(yellow, top left) from \citet{cepheid_lmc}, Long-period semi-regular
variables (SRVs, orange, top center) and Mira stars in the LMC (red,
top right) from \citet{lpv_lmc}, LMC eclipsing binaries (green, middle
left) from \citet{eclipsing_lmc}, the SMC eclipsing binaries (green,
middle center) from \citet{eclipsing_smc}, LMC RR Lyrae (blue, bottom
left) and SMC RR Lyrae (purple,
bottom center) from \citet{rrl_smc}. Also shown are the QSO and AGN
(black, middle right) from \citet{qso_agn}. It helps enormously that
all of the stellar variables above are located at approximately the
same heliocentric distance. Therefore, for many of these, the $G$
magnitude distribution is simplified, as illustrated by the clustering
of the LMC Cepheids and LPVs. The clustering is most pronounced for
the RR Lyrae: for these pulsators, the amplitude-luminosity relation
induces only a modest change in the apparent magnitude.

\begin{figure*}
  \centering
  \includegraphics[width=0.98\textwidth]{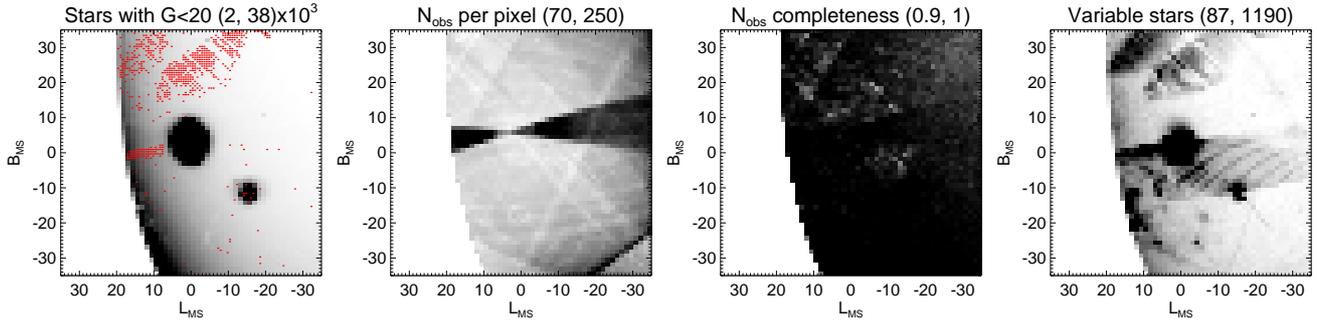}
  \caption[]{\small Statistics of the Gaia DR1 observations of the
    Magellanic system. This shows the $70^{\circ}\times70^{\circ}$
    area centered on the LMC in the Magellanic Stream coordinate
    system. Density maps have pixels with 1.25 degree on a side. The
    pixel values corresponding to black and white are given in
    brackets in the title of each panel. {\it First panel:} Number
    density of stars with $G<20$. Small red dots mark the pixels
    identified as strongly affected by the cross-match failures and
    excluded from the subsequent analysis. {\it Second panel:} Mean
    number of observations per pixel on the sky. Note the dark region
    corresponding to the Ecliptic Pole scanning.  {\it Third panel:}
    Completeness due to $N_{obs}>70$ cut. {\it Fourth panel:} Density
    of ``variable'' sources using the cut shown in
    Figure~\ref{fig:selection}. Note that apart from the LMC and the
    SMC, a number of over-dense regions appear. These are the portions
    of the Gaia DR1 sky most affected by cross-match failures.}
   \label{fig:selection2}
\end{figure*}

Motivated by the tight bunching of the previously identified
Magellanic RR Lyrae in the $(G, \Amp)$ space, we propose a simple
selection box shown in the right panel of the bottom row of
Figure~\ref{fig:variables}. Sources above the diagonal line are
predicted to exhibit variability larger than expected for a constant
star at the given $G$ magnitude. Note that this is a conservative
variability threshold and most non-variable sources have much lower
$\Amp$ values (at given magnitude).  However, we believe this choice
is warranted given the GSDR1 teething problems with the source
cross-match (see Section~\ref{sec:problems} for details). The diagonal
line slices through the cloud of RR Lyrae approximately where the RRab
and the RRc pulsators separate (for the LMC, these are indicated with
different shades of blue). Therefore, our RR Lyrae sample consists
almost entirely of the RR Lyrae of the ab type. The vertical
boundaries are chosen to include both the LMC and the SMC RR
Lyrae. Note that the apparent magnitude of the RR Lyrae in the LMC is
offset $\sim 0.5$ magnitude brighter compared to that of the SMC. This
reflects the difference in the line-of-sight distance to the Clouds:
the LMC is at $49.97$ kpc \citep[see][]{LMCdist} and the SMC is at
$62.1$ kpc \citep[see][]{SMCdist}. Converted into distance moduli,
these correspond to 18.509 and 18.965 for the LMC and the SMC
respectively. It is clear from the Figure that the selection proposed
is neither complete nor pure. The objects chosen using this simple
boundary will not be limited to the Magellanic RR Lyrae exclusively:
some of the Magellanic eclipsing binaries will also be
included. Additionally, a small number of variable QSO and AGN can
pass these variability cuts too. We discuss the sample's completeness
and purity in Section~\ref{sec:selection}.

\subsection{Selection biases, galaxies and artifacts}
\label{sec:problems}

Having glanced at the distribution of genuine variable stars in the
plane of $(G, \Amp)$, let us inspect the behavior of the bulk of the
Gaia sources in and around the Clouds. Figure~\ref{fig:selection}
presents (the logarithm of) the density of sources in the $(G, \Amp)$
space for a $70^{\circ}\times70^{\circ}$ region centered on
$(\LMS, \BMS)=(0^{\circ}, 0^{\circ})$. More precisely, the first
panel gives the view of the foreground/background as the Clouds
themselves are excised from this picture, while the second and the
third panels display stars in the LMC and the SMC respectively. As
predicted above, most stars lie well below the diagonal line
segregating variable and non-variable objects. Additionally, in the
leftmost panel, very few stars enter the Magellanic RR Lyrae box in
the top right corner of the plot. The second and third panels confirm
that this box is populated with RR Lyrae stars, whose magnitude
distributions are offset with respect to each other due to the the
difference in the heliocentric distances of the Clouds as discussed in
Section~\ref{sec:variables}.

\begin{figure*}
  \centering
  \includegraphics[width=0.98\textwidth]{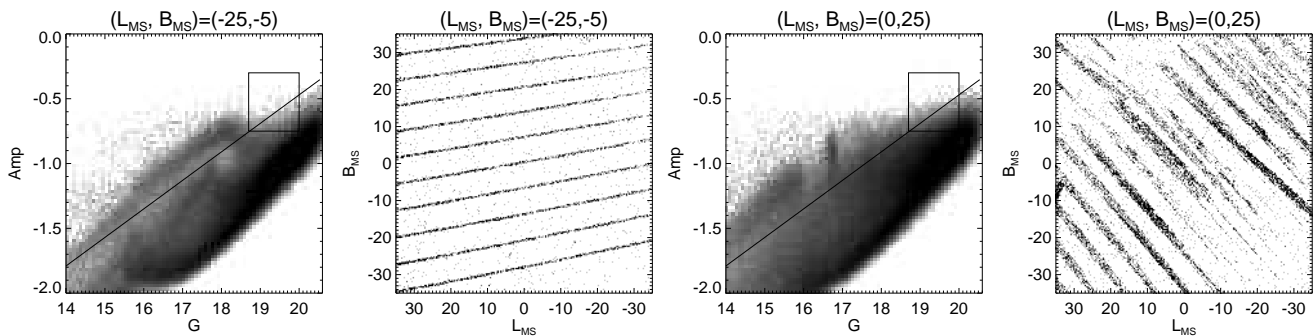}
  \caption[]{\small Examples of the \Gaia\ cross-match failures in
    selected $10^\circ \times 10^\circ$ areas around the Magellanic Clouds. {\it First and Third
      Panels:} Density of stars in the plane of \Amp and $G$
    magnitude. Note the sharp diagonal features symptomatic of flux
    mis-allocation. Variable stars are required to lie above the black
    diagonal line, while the Magellanic RR Lyrae must also be within
    the black box shown in top right corner. {\it Second and Fourth
      Panels:} On-sky density distribution of ``variable'' stars. The
    \Gaia\ scanning pattern is clearly visible, thus emphasizing the
    spurious nature of many of the stars identified as ``variable'' in
    these regions.}
   \label{fig:problem}
\end{figure*}

Apart from the many expected features, the distributions shown in the
first three panels of Figure~\ref{fig:selection} also reveal several
odd-looking sub-structures, many of which run diagonally across the
$(G, \Amp)$ plane over a wide range of magnitudes. We believe that
most of these sharp over-densities in the variability-magnitude space
are spurious, and are caused by cross-match failures in the
GSDR1. This is confirmed in the rightmost panel of
Figure~\ref{fig:selection2} where the on-sky density distribution of
all nominally variable objects (i.e. stars above the black diagonal
line) is displayed. Apart from the obvious over-densities clearly
associated with the LMC and the SMC, there are many regions with
sharply defined boundaries with a strong excess of ``variable''
objects. Figure~\ref{fig:problem} provides further insight into the
nature of these artifacts. The Figure zooms in onto several over-dense
regions visible in the right panel of Figure~\ref{fig:selection2} and
shows that these over-densities resolve into thin, mostly well-aligned
strips. For the $10^{\circ}\times10^{\circ}$ region centered on
$(\LMS, \BMS)=(-25^{\circ}, -5^{\circ})$ (shown in the first and
second panels of Figure~\ref{fig:problem}), the strips are less than a
degree wide (in fact, their cross-section approximately matches the
\Gaia's field of view size of $0.65^{\circ}$) and the separation
between the strips appears constant and equal to
$\sim5^{\circ}$. Given the tight alignment between the strips, it
seems likely that the problem occurred over a small range of
epochs. Given the sharp diagonal sequence sitting above the variable
selection line and turning over at $G\sim 18$ (see left panel of the
Figure), it appears that stars over a wide range of magnitudes were
affected. Based on the diagnostics presented in
Figures~\ref{fig:selection2} and ~\ref{fig:problem}, we conjecture
that a fault in the object cross-match procedure is the cause of these
spurious features. At a small number of epochs (as indicated by the
sparseness of the strips), stars were assigned flux from unrelated
objects, thus making the otherwise non-variable sources appear
``variable''.

In the presence of these striking artifacts, it is comforting to see
that very few spurious objects seem to have entered the designated RR
Lyrae box. Bear in mind, however, that the exact pattern of spurious
features changes from location to location as displayed in the third
and fourth panels of Figure~\ref{fig:problem} where stars from the
region centered on $(\LMS, \BMS)=(0^{\circ}, 25^{\circ})$ are
shown. Here, the width and the distance between the strips seems to be
variable, indicating that the cross-match has likely faltered for
objects observed at several epochs (or ranges of epochs). In the
amplitude-magnitude space, the familiar diagonal feature is visible,
although it seems to be less pronounced at $G>18$. Nonetheless, the RR
Lyrae box appears to be slightly more contaminated compared to the
levels seen for the stars in the $(\LMS, \BMS)=(-25^{\circ},
-5^{\circ})$ region.

If the problems with the GSDR1 source cross-match are the cause of the
spurious variability discussed above, then the objects affected ought
to exhibit abnormal astrometric behavior as well. This seems indeed to
be the case as illustrated in the rightmost panel of
Figure~\ref{fig:selection}. Here, (the logarithm of) the astrometric
excess noise is shown as a function of the variability amplitude
\Amp\ for stars with magnitudes consistent with the Magellanic RR
Lyrae. The objects appear to sit in two separate clusters in this 2D
plane: the one that stretches upward from low to high ${\rm AEN}$
values, and the one which seems to be composed only of objects with
high ${\rm AEN}$. By examining the catalogues of the \Gaia\ sources
observed by the SDSS, it has become clear that the latter (the
isolated cloud of high ${\rm AEN}$ objects) mostly consists of
galaxies (or, perhaps, their central compact and high
surface-brightness parts). Note that the larger sequence, sitting
below the galaxy cloud, appears to change its shape as a function of
\Amp: in other words, there is a noticeable correlation between the
photometric variability and poor astrometric fit, especially for
objects with $\log_{10}({\rm AEN})>0.2$. Therefore, we choose to cull
the contaminating galaxies as well as the objects most affected by
cross-match failures by requiring $\log_{10}({\rm AEN})$ to be lower
than a certain threshold value, the choice for which is discussed
below.

A distinctive feature of the \Gaia\ mission is the non-uniformity of
the sky coverage. The \Gaia's scanning law produces strong patterns on
the celestial sphere in terms of the numbers of visits per
location. At the time of the GSDR1 release, some corners of the Galaxy
barely had 10 \Gaia\ observations. The number of visits not only
determines the overall depth of the $G$-band photometry, but also
controls the significance of the source's variability. Moreover, the
variability statistic \Amp\ will evolve as the number of observations
grows, depending on the shape of the lightcurve and the period of the
star. The second panel of Figure~\ref{fig:selection2} shows the
average number of CCD observations per pixel on the sky. The strongest
feature is the ecliptic scan region which was repeatedly imaged by
\Gaia\ at the beginning of the mission. The map shows changes in
$N_{\rm obs}$ per source across the sky, i.e. the number of individual
CCD transits. Given that the \Gaia's focal plane contains an array of
9 CCDs, this number must be divided by 9 to get an approximate number
of visits of the given object. The number of visits is likely lower
for fainter stars as they may not be detected in every FoV transit and
it is more likely that, due to the priority given to brighter objects,
they may not be allocated a window. As the Figure demonstrates, while
the variation in the number of observations is markedly apparent, most
stars around the Magellanic Clouds have traversed the \Gaia's focal
plane at least 8 times ($N_{\rm obs}>70$). Indeed, as the third panel
demonstrates, requesting the minimal of $N_{rm obs}=70$ induces only
minor incompleteness, which can easily be corrected for.

\subsection{Magellanic RR Lyrae sample}
\label{sec:selection}

Guided by the GSDR1 properties of known variable
stars as well as the behavior of the data as a function of the number
of observations and the resilience of the variability statistic
against the artifacts induced by the failures of the cross-match
procedure, we put forward the following selection cuts aimed to
produce a sample of RR Lyrae candidates around the LMC and the SMC.

\begin{align}
  \label{eqn:selection}
  \begin{split}
    \Amp > 0.22G-4.87 & ~~~i \\
    \left.
    \begin{aligned}
      \log_{10}\left({\rm AEN}\right)<0.2,  ~~~{\rm weak} \\
      \log_{10}\left({\rm AEN}\right)<-0.2,  ~~~{\rm strict}
    \end{aligned}  \right \} & ~~~ii\\
    18.7 < G < 20.0 & ~~~iii
    \\
    N_{\rm obs}>70 & ~~~iv
    \\
    E(B-V)<0.25 & ~~~v \\
    \left.
    \begin{aligned}
    -0.75 < \Amp < -0.3,  ~~~{\rm weak}  \\
    -0.65 < \Amp < -0.3,  ~~~{\rm strict}
        \end{aligned}  \right \} & ~~~vi\\
    b<-15^{\circ} & ~~~vii
  \end{split}
\end{align}
\begin{figure}
  \centering
  \includegraphics[width=0.48\textwidth]{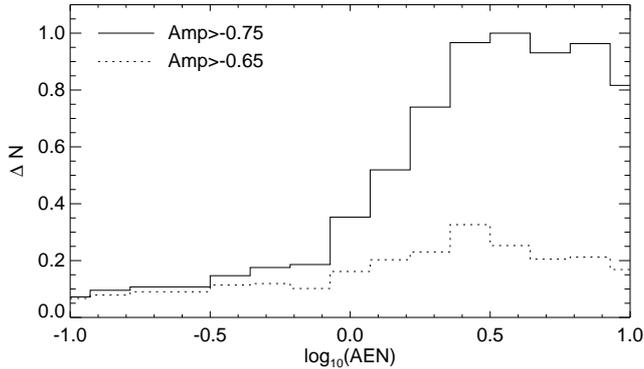}
  \caption[]{\small Excess of spurious variable stars as a function of
    the $\AEN$ threshold for two different variability amplitude
    cuts. This shows the difference in the number of stars in an area
    around $(\LMS, \BMS)=(\-25^{\circ},-5^{\circ})$ (see
    Figure~\ref{fig:problem}) compared to an area largely unaffected
    by cross-match breakdown, namely $(\LMS,
    \BMS)=(\-25^{\circ},-25^{\circ})$. If $\Amp>-0.75$ cut is imposed,
    then a cut of $\AEN<0.2$ gets rid of $\sim50\%$ of the spurious
    variables, while $\AEN<-0.2$ only leaves $\sim15\%$ artifact
    contamination. Note that a similar level of $\sim15\%$ can be
    achieved with $\Amp >-0.65$ and $\AEN<0.3$ cuts.}
   \label{fig:problem_excess}
\end{figure}

\noindent The first cut selects the likely variable objects; the
second one gets rid of galaxies and the objects most affected by
cross-match failures - this cut be made stronger if a cleaner sample
of RR Lyrae is required; the third one limits the magnitude range to
that populated by the LMC and the SMC RR Lyrae; the fourth requires at
least 8 visits to the given location; the fifth eliminates the areas
most affected by the Galactic dust (note that this cut is only applied
outside of a 4 degree radius from the LMC's centre); the sixth cut
limits the overall variability amplitude; finally, the seventh cut
gets rid of the fields too close to the Galaxy's disc. Additionally,
there are two areas in the vicinity of the LMC that are affected by
the presence of spurious variables more than others. This is i) an
area with $15^{\circ} < \LMS < 5^{\circ}$ visible as a dark thin bar
to the left of the LMC in the rightmost panel of
Figure~\ref{fig:selection2} and ii) the area around $(\LMS,
\BMS)\sim(-5^{\circ},25^{\circ})$ with a pattern of artifacts
displayed in the third and fourth panels of
Figure~\ref{fig:problem} \footnote{This area is only few degrees away
  from the second brightest star on the sky, Canopus, and therefore
  may have been affected by the star's ghost images.}. We eliminate a
small number of the most affected pixels in these two areas as
follows. Given that very few genuine variable stars exhibit
variability levels higher than $\Amp=-0.4$ at $G>19$, we create a map
of number counts of stars with $-0.37<\Amp<0.5$ and $19<G<20.5$ and
cull all pixels with values above the 95th percentile. As shown in
third panel of Figure~\ref{fig:problem}, the second most affected area
has a sharp feature in the $(G, \Amp)$ space at $G\sim17$. Therefore,
we build a map of number counts of stars with $-1<\Amp<-0.8$ and $16.7
< G<16.9$ and get rid of pixels with values above the 95th
percentile. All of the affected pixels are marked with red dots in the
left panel of Figure~\ref{fig:selection2}.

\begin{figure*}
  \centering
  \includegraphics[width=0.98\textwidth]{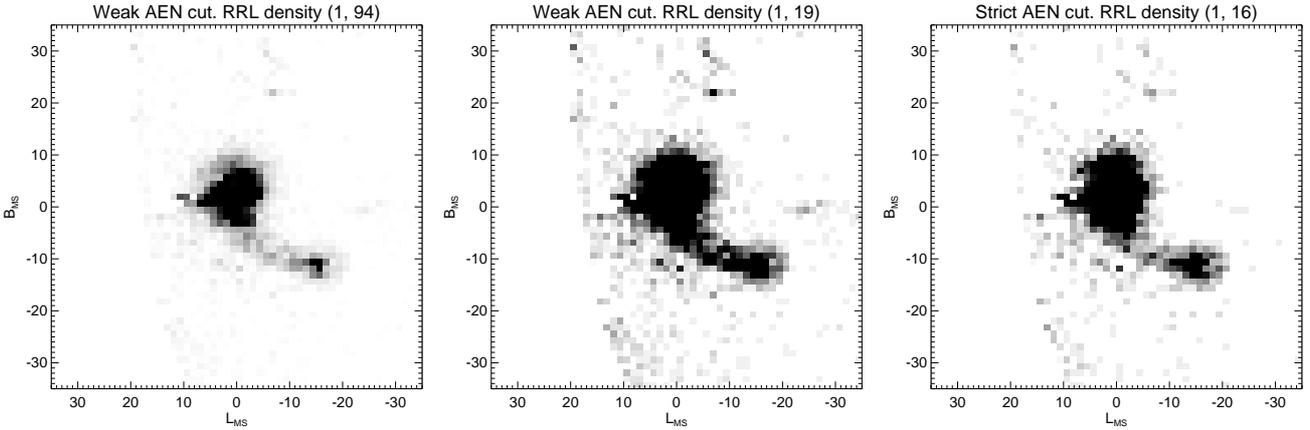}
  \caption[]{\small Density of the RR Lyrae candidate stars in
    $70^{\circ}\times70^{\circ}$ area centered on the LMC in the
    Magellanic Stream coordinate system. Pixels are 1.25 degree on a
    side. The pixel values corresponding to black and white are given
    in brackets in the title of each panel. {\it Left and Center:}
    $\sim$23,000 stars selected using (amongst others) a $\log_{10}({\rm
      AEN})<0.25$ cut (see Figure~\ref{fig:selection} for
    details). The difference between the two panels is only in the
    maps' dynamic range as indicated in the panel titles. Both Clouds
    are clearly visible as well as a bridge connecting them, with a
    cross-section roughly matching that of the SMC. {\it Right:}
    Density map of the 10501 RR Lyrae candidates selected using a
    stricter $\log_{10}({\rm AEN})<-0.2$ cut. Here, most of the
    artifacts related to the cross-match visible in the center panel
    disappear, albeit at the expense of the noticeable reduction in
    the sample size.}
   \label{fig:bridge}
\end{figure*}
\begin{figure*}
  \centering
  \includegraphics[width=0.98\textwidth]{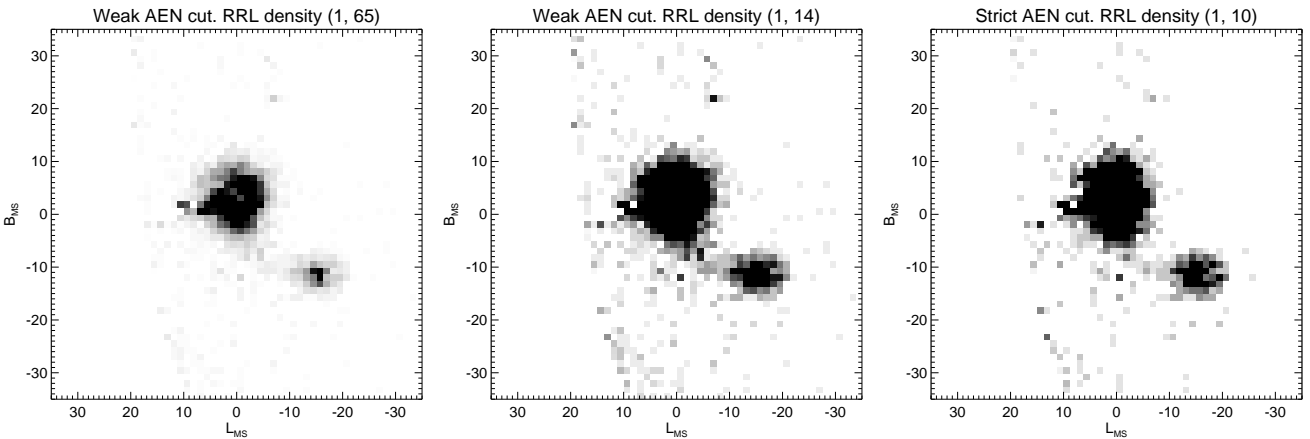}
  \caption[]{\small Same as Figure~\ref{fig:bridge} but with
    an $\Amp>-0.65$ cut.}
   \label{fig:bridge2}
\end{figure*}

The combination of the first and the last cut in
Equation~\ref{eqn:selection} yields $\sim3.9 \times 10^6$ objects in
the $80^{\circ}\times80^{\circ}$ area around the LMC. If in addition a
weak (strict) cut on ${\rm AEN}$ is imposed, the number of variable
objects shrinks to $\sim2.8 \times 10^6$ ($\sim1.6 \times 10^6$)
sources. The sample shrinks drastically if these two criteria are
applied in combination with the magnitude cut, leaving a total of
$67,000$ likely variable objects with magnitudes consistent with
Magellanic RR Lyrae. The application of all cuts in
Equation~\ref{eqn:selection} as well as the masking of bad pixels
described above, produces the final sample of $\sim 21,500$ RR Lyrae
candidates. These numbers are consistent with the expectation for the
total number of RR Lyrae around the Clouds. For example,
\citet{ogle_rrl} report some $45,000$ RR Lyrae found as part of the
OGLE-IV Magellanic campaign. Our sample is smaller even though the
area covered is significantly larger. This is because the completeness
of our selection is far from $100\%$ as indicated by the diagonal line
slicing right through the clusters of RR Lyrae in
Figure~\ref{fig:variables}. Additionally, given that this line passes
through the SMC RR Lyrae at higher values of \Amp, the completeness of
the SMC RR Lyrae sample is expected to be lower than that of the
LMC. We estimate the completeness of our selection by counting the
number of the previously identified RR Lyrae stars recovered around
the LMC and the SMC. Namely, we detect $\sim38\%$ of the LMC RR Lyrae
reported as part of the GDR1 \citep[see][for details]{rrl_lmc} and
$\sim12\%$ of the SMC RR Lyrae discovered by \citet{rrl_smc}. The
above numbers are for the ``weak'' $\AEN<0.2$ cut. If a ``strict''
$\AEN<-0.2$ cut is applied, the completeness drops to $\sim 13\%$ and
$\sim8\%$ respectively. As shown below, an alternative RR Lyrae
selection can be used, where the amplitude cut is tightened to
$\Amp>-0.65$ while keeping the weak cut on astrometric excess noise
($\AEN<0.3$): the completeness of this selection is $\sim 30\%$ and
$\sim11\%$ for the LMC and the SMC RR Lyrae correspondingly. Finally,
if strict cuts are used for both variability and amplitude, the
completeness is minimal at the level of $<10\%$ for both the LMC and
the SMC RR Lyrae.

The contamination of the GSDR1 RR Lyrae sample can be gauged by
counting the number of stars classified as RR Lyrae candidates using
GSDR1 information only, but not by other variability surveys. This
procedure can only be implemented in the vicinity of the LMC and the
SMC where published RR Lyrae datasets exist. Using the samples
presented by \citet{rrl_lmc} and \citet{rrl_smc}, the contamination of
the RR Lyrae sample analysed here is between $30\%$ and $40\%$. This
is much worse than is typically achieved by targeted RR Lyrae searches
\citep[see e.g.][]{drake_rrl,gabriel_rrl,nina_rrl}. Nonetheless, the
purity of our RR Lyrae selection is higher than that of samples of
distant BHB stars assembled using deep broadband photometry with
surveys such as SDSS and DES \citep[see
  e.g.][]{coldveil,void,sgrprecession,lmc_streams}.

The GSDR1 RR Lyrae sample purity as estimated above does not vary
dramatically as the \Amp\ and AEN cuts are changed from weaker to
stronger. However, the excess of spurious variable stars in areas
affected by cross-match failures can be reduced significantly by
dialing the variability amplitude and the astrometric excess noise
thresholds. This is illustrated in
Figure~\ref{fig:problem_excess}. Here, the excess of RR Lyrae
candidate stars in the problematic area centered on $(\LMS,
\BMS)=(-25^{\circ}, -5^{\circ})$ with respect to the count in a
relatively un-affected area around $(\LMS, \BMS)=(-25^{\circ},
-25^{\circ})$ is shown as a function of the AEN cut for two different
$\Amp$ choices. As shown in the Figure, with $\Amp>-0.75$, the
weak cut $\AEN<-0.2$ gets rid of $\sim50\%$ of the spurious excess
(black solid line). Making the AEN criterion stricter,
i.e. $\AEN<-0.2$ leaves only $<20\%$ contamination. On the other hand,
similar purity in this area can be achieved if the variability
amplitude threshold is higher at $\Amp>-0.65$ and $\AEN<0.3$. The two
choices for the combination of the $\Amp$ and the $\AEN$ cuts deliver
similar levels of purity in the cross-match affected areas, albeit the
latter yields a higher completeness (as described above). In what
follows, we use different combinations of $\Amp$ and the $\AEN$
thresholds and explore how the properties of the outer environs of the
LMC and the SMC change as the completeness and the purity of GSDR1
sample of RR Lyrae evolves.

\section{The Magellanic bridges}
\label{sec:bridges}

\subsection{The RR Lyrae bridge}
\label{sec:rrl_bridge}

\begin{figure}
  \centering
  \includegraphics[width=0.45\textwidth]{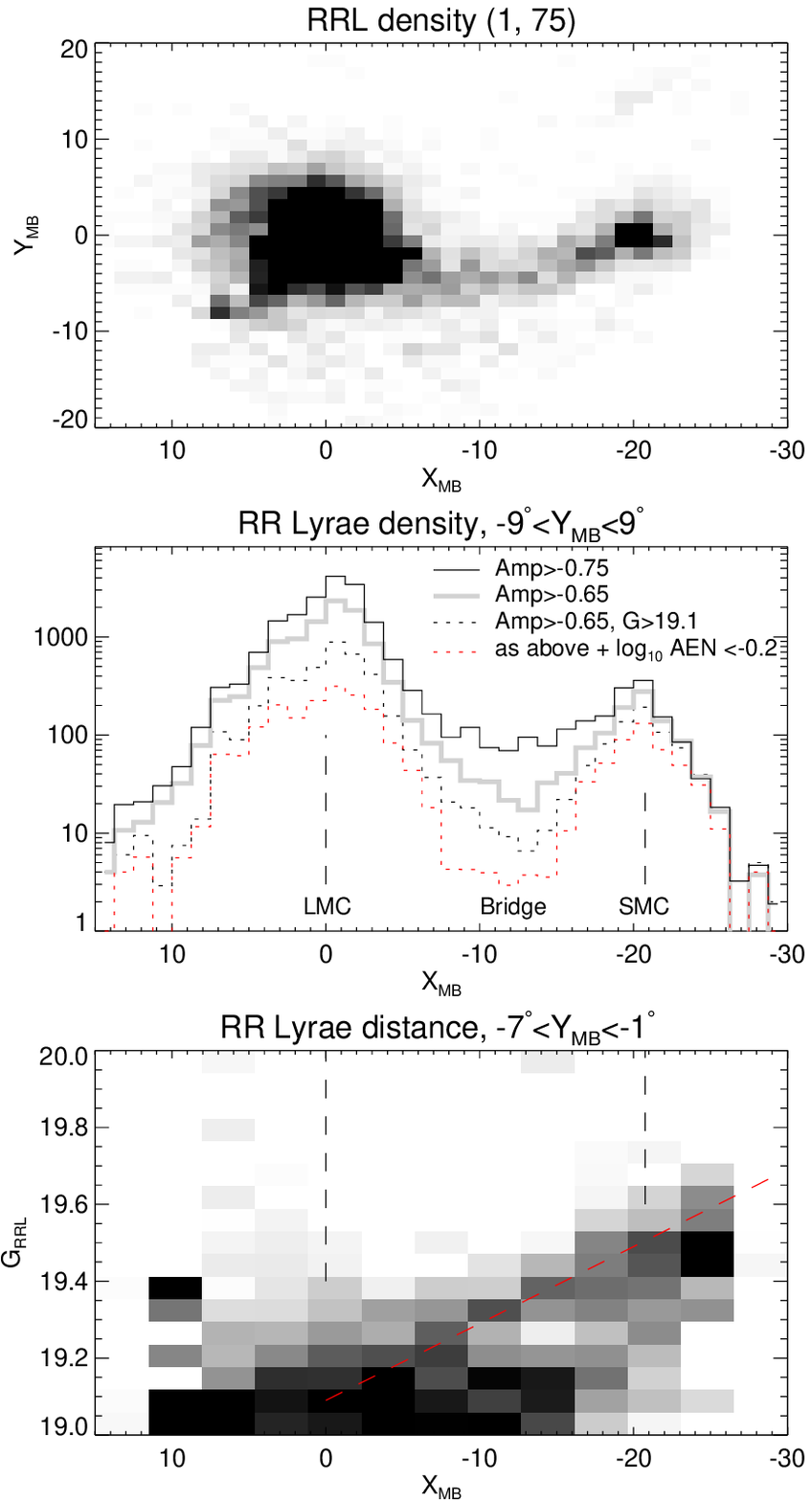}
  \caption[]{\small {\it Top:} Density of the Gaia RR Lyrae candidates
    in the Magellanic Bridge (MB) coordinate system in which both the
    LMC and the SMC lie on the equator. The pixel size is 1.25 degrees
    on a side. {\it Middle:} Density of the RR Lyrae candidate stars
    along the MB equator. A foreground/background model (linear along
    MB latitude $\BMB$ and different for each $\LMB$) is
    subtracted. The inset explains the cuts applied for each of the
    histograms. {\it Bottom:} Variation of the apparent magnitude of
    the RR Lyrae along the bridge. For these RR Lyrae, a stricter cut
    on variability amplitude was applied, namely ${\rm
      Amp}>-0.65$. Between the LMC and the SMC, two structures at
    distinct distances are visible: one at the distance of the LMC,
    i.e. at $G\sim 19$ and one connecting the LMC and the SMC (at
    $G\sim 19.5$). Red dashed line gives the approximate behavior
    $G=19.02-0.2\LMB$ of the more distant of the two RR Lyrae
    structures.}
   \label{fig:aligned}
\end{figure}

Figure~\ref{fig:bridge} shows the density of the GSDR1 RR Lyrae
candidate stars in the MS coordinate system. These are selected using
the criteria presented in Equation~\ref{eqn:selection}, in particular
by applying the weak cut on variability. The left and center panels
differ only in the dynamic range of the pixel values: the density map
shown on the left saturates at high values while the map in the center
saturates at much lower density levels. Traced with RR Lyrae, the
Clouds do not appear very round. In the MS coordinate system, the LMC
is stretched in the North-South direction, while the SMC seems to be
vertically squashed. This stretching can also be seen in
Figure~\ref{fig:starcounts} which shows the raw star counts in the
Magellanic Bridge coordinates.

\begin{figure*}
  \centering
  \includegraphics[width=0.98\textwidth]{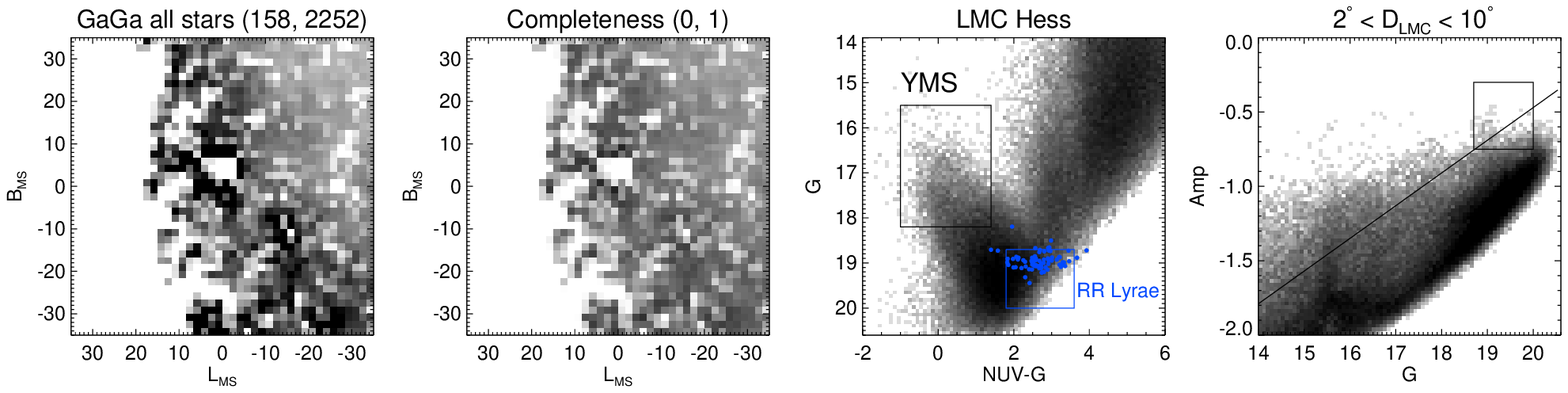}
  \caption[]{\small Selecting Magellanic RR Lyrae and Young Main
    Sequence stars with GaGa = \Gaia+{\it Galex}. {\it First panel:}
    Density map of stars with both \Gaia\ and {\it Galex}
    detections. The pixel size is 1.667 degrees on a side. {\it Second
      panel:} Completeness map of GaGa. {\it Third panel:} Logarithm
    of density in the color-magnitude space (Hess diagram) spanned by
    $G$ and $NUV-G$ for stars with $2^{\circ}<D_{LMC}<10^{\circ}$. Two
    selection boxes are shown: one aimed at identifying Magellanic RR
    Lyrae stars (blue) and one for Young Main Sequence stars
    (black). {\it Fourth panel:} Logarithm of density of GaGa sources
    in the space spanned by \Amp\ and $G$ for a region around the
    LMC. The LMC RR Lyrae are clearly identified in the designated
    selection box. Note that we use a slightly different cut on
    astrometric excess noise, i.e. $\log_{10}({\rm AEN})<-0.1$}
   \label{fig:gaga_selection}
\end{figure*}

In both panels, a narrow and long structure linking the SMC and the
LMC is obvious. This ``bridge'' connects the Eastern side of the SMC
and the Southern edge of the LMC. Its width roughly matches the extent
of the SMC. The right panel of the Figure shows the density of RR
Lyrae candidates selected using a stricter cut on AEN. While the
bridge is clearly less prominent in the right panel, its width and
length remain largely unchanged. A version of the RR Lyrae density map
is shown in Figure~\ref{fig:bridge2}. Here, the cut on the variability
amplitude is stricter, i.e. $\Amp>-0.65$, which allows us to relax the
cut on astrometric excess noise, i.e. $\AEN<0.3$ (left and center
panels, also see Section~\ref{sec:rrl} for the discussion of the
effects of different selection criteria). Finally, the right panel of
the Figure shows a density map of the ``double-distilled'' sample of
RR Lyrae candidates: with $\Amp>-0.65$ and $\AEN<-0.2$. Reassuringly,
the bridge remains visible, regardless of the level of ``cleaning''
applied. However, the number of stars in the bridge drops
significantly with stricter cuts. Importantly, as the completeness and
the purity varies, across all 6 panels of the Figures~\ref{fig:bridge}
and \ref{fig:bridge2} combined, the shape of the GSDR1 RR Lyrae
distribution looks consistent.

Comparing the RR Lyrae density maps to the distribution of artifacts
shown in the rightmost panel of Figure~\ref{fig:selection2}, we note
that the bridge does not appear to follow any particular spurious
over-density and its borders are not coincident with boundaries of the
cross-match affected areas. However, the pronounced decrease in the
bridge number counts on moving from the middle panel of
Figure~\ref{fig:bridge} to the right panel of Figure~\ref{fig:bridge2}
may nonetheless imply that while the shape of the bridge is robust,
its density levels are affected by spurious variables. Comparing to
the bottom panel of Figure~\ref{fig:ebv}, it is clear that none of the
features in the RR Lyrae density map are coincident with the details
of the Galactic dust distribution either. Therefore, we judge the RR
Lyrae map not to be seriously affected by the effects of interstellar
extinction. Overall, we conclude that the RR Lyrae bridge seen between
the two Clouds is a genuine stellar structure.

We investigate the properties of the GSDR1 RR Lyrae bridge using the
Magellanic Bridge coordinate system defined above. In these
coordinates, the RR Lyrae bridge runs parallel to the equator and is
limited to $-7^{\circ}<\BMB<-1^{\circ}$. To obtain the centre and the
width of the RR Lyrae distribution in each bin of $\LMB$, we fit a
model which includes a linear foreground/background and a Gaussian for
the stream's signal. The measured centroid and the width values are
reported in Table~\ref{tab:rrl}. The top panel of
Figure~\ref{fig:aligned} gives the density of RR Lyrae selected using
the ``weak'' version of the cuts presented in
Equation~\ref{eqn:selection}. The middle panel of the Figure shows the
background-subtracted number density profile along the bridge (black
histogram). For comparison, grey (red) histogram shows the number
density profile obtained using the sample of GSDR1 RR Lyrae obtained
with $\Amp>-0.65$ and $\AEN<0.3$ ($\AEN<-0.2$) cuts. Regardless of
which set of the RR Lyrae selection criteria is used, the bridge
density profile appears to have a depletion around the mid-point,
i.e. at $\LMB\sim -12^{\circ}$. The simplest interpretation of this
behavior is that the objects in this area of the sky come from two
groups of stars, one around the LMC and one emanating from the SMC,
each with a negative density gradient away from each Cloud. This could
also explain the change in the curvature of the bridge at around
$\LMB\sim-10^{\circ}$.

Given that the LMC and the SMC are offset with respect to each other
along the line of sight, it should be possible to test the above
idea. To that end, the lower panel of Figure~\ref{fig:aligned} shows
the apparent magnitude distribution of the $\Amp>-0.65$ and $\AEN<0.3$
RR Lyrae with $-7^{\circ}<\BMB<-1^{\circ}$ as a function of the MB
longitude $\LMB$. As expected, the bulk of the LMC's RR Lyrae are
around $G\sim19$ while those belonging to the SMC aggregate in the
vicinity of $(\LMB, G)=(-20.75^{\circ}, 19.5)$. Between the LMC and
the SMC, i.e. $\LMB=0^{\circ}$ and $\LMB=-20.75^{\circ}$, there appear
to be two distinct sequences. First, the more pronounced, at constant
$G=19$ extends from $\LMB=0^{\circ}$ to at least
$\LMB\sim-15^{\circ}$, or possibly further. Additionally, there is a
clear second, albeit seemingly less populated, sequence which appears
to connect the SMC and the LMC. Therefore, at a number of locations
along the bridge there are two stellar over-densities, one at the
distance of the LMC, and one traveling from the SMC towards the
LMC. In the Figure, the debris around the LMC's nominal distance
appear to be more numerous at each sight-line through the
bridge. However, this apparent line-of-sight distribution is
misleading as the RR Lyrae sample completeness is a strong function of
the $G$ magnitude. Given that at the SMC's distance, the completeness
is at least 3 times lower, it is entirely possible that the bridge
contains as much distant (i.e. at distances between the LMC and the
SMC) debris as there is at the LMC's distance. In the future (and
certainly with \Gaia\ DR2), it should be possible to disentangle the
bridge debris in 3D. However, already with the current data it seems
likely that the inflection point in the bridge centroid at around
$\LMB=-10^{\circ}$ is due to the change in the ratio of the debris
groups at different distances.

\begin{figure*}
  \centering
  \includegraphics[width=0.98\textwidth]{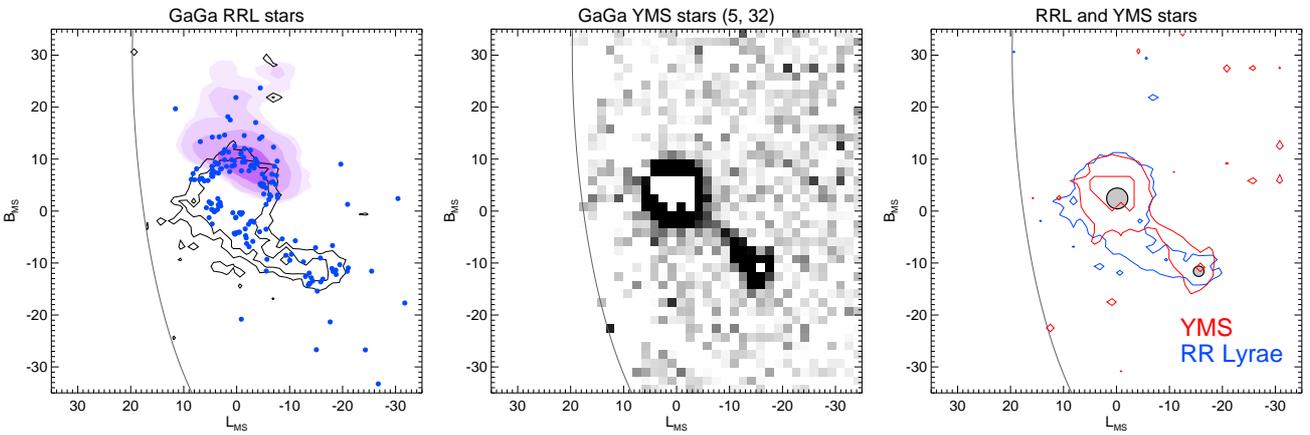}
  \caption[]{\small {\it Left:} Positions of 167 GaGa RR Lyrae
    candidates. Contours show the density of Gaia DR1 RR Lyrae
    candidates. Outside of the main LMC body, there are two clear
    overdensities of the GaGa RR Lyrae. First, the one between the LMC
    and the SMC, corresponding to the bridge reported in
    Figure~\ref{fig:bridge}. Additionally, there is a plume of GaGa RR
    Lyrae which extends to the North of the LMC in agreement with BHB
    detections (shown as purple density contours) reported in
    \citet{lmc_streams}. {\it Center:} Density of $\sim$45,000 YMS
    candidates in GaGa. Note that the YMS stars trace a different,
    much narrower bridge, clearly offset from the RR Lyrae tail. {\it
      Right:} Comparison of the density distributions of the Gaia RR
    Lyrae candidates (blue) and GaGa YMS candidates (red).}
   \label{fig:gaga_bridge}
\end{figure*}

Using the background-subtracted density profile discussed above, it is
feasible to estimate the total number of RR Lyrae in the
bridge. However, because the GSDR1 RR Lyrae completeness is a strong
function of apparent magnitude, the complex 3D structure of the bridge
also needs to be taken into account. We define the bridge extent as
that limited by $-15^{\circ}<\LMB<-10^{\circ}$. This is motivated by
the line-of-sight map shown in the lower panel of
Figure~\ref{fig:aligned}. Outside of this $\LMB$ range, the RR Lyrae
sample is dominated by stars that are currently still (likely) part of
the LMC or the SMC. For the calculation below, we assume that the LMC
provided the bulk of stars with $G<19.1$ and the stars with $G>19.1$
are mostly from the SMC (note, however, the discussion below and in
Section~\ref{sec:sims}). There are 75 RR Lyrae with $18.7 < G < 19.1$,
which given the $\sim30\%$ completeness (see
Section~\ref{sec:selection}) would translate into 250 RR Lyrae stars
from the LMC - or at distances consistent with that of the LMC - in
this region. There are 40 RR Lyrae with $G>19.1$, all of which we
tentatively assign to the SMC. As the dashed black histogram in the
middle panel of Figure~\ref{fig:aligned} demonstrates, the density of
these stars as a function of $\LMB$ is reasonably flat within the
bridge range specified above. Assuming the variation in completeness
from 0.3 at $G=19$ to 0.11 at $19.5$ and assuming the distance to the
SMC tidal tail goes like $G=19.02-0.2\LMB$ (shown as red dashed
diagonal line in the bottom panel of Figure~\ref{fig:aligned}), we
estimate a total of 240 RR Lyrae that could have been pulled out from
the SMC. Note that the number of the RR Lyrae with $G>19.1$ detected
in this area drops to 14 if a strict cut on astrometric excess noise
($\AEN<-0.2$) is applied. This would imply that at most 70 RR Lyrae
may exist here. Worse still, we have not corrected any of these
numbers for contamination, which we assumed to be (at least
approximately) taken care of by the subtraction of the background
model. Of course, if this region is overdense in spurious variables,
the contamination will be far from zero. Also note that the above
differentiation of the RR Lyrae into those belonging to the LMC and
the SMC solely based on their apparent magnitude is very
simplistic. This classification should be carried out using the actual
distances to these stars.

The above discussion also glosses over some important details of the
LMC's structure (such as line-of-sight distance gradients) as well as
the details of its interaction with the SMC (i.e. the stars consistent
with the LMC's distance could in fact be the SMC debris stripped much
earlier). The latter will be dealt with in
Section~\ref{sec:sims}. With regards to the former, let us point out
that a pronounced distance gradient has been measured across the LMC's
disc \citep[see e.g.][]{Mackey2016}. This gradient is positive in the
direction of the decreasing $\LMB$. This necessarily implies that at
least some of the debris with $G>19.1$, are in fact part of the
LMC. It is not clear, however, how far this LMC population can
stretch. \citet{Saha2010} find evidence fir the LMC stars at angular
separations of $\sim15^{\circ}$. According to \citet{Mackey2016}, on
the other side of the dwarf, the disc is perturbed into a stream-like
structure visible at $\LMB\sim13.5^{\circ}$. If a counterpart to the
\citet{Mackey2016} ``stream'' exists, then many of the distant RR
Lyrae stars in the bridge at $\LMB\sim-10^{\circ}$ (and maybe as far
as $\LMB\sim-15^{\circ}$) are from the LMC's disc. We investigate this
possibility further with simulations in Section~\ref{sec:sims}. This
leaves the nature of the portion of the RR Lyrae with constant
$G\sim19$ and $10^{\circ}>\LMB>-15^{\circ}$, seen as a dark horizontal
bar in the bottom panel of Figure~\ref{fig:aligned} rather
unclear. Curiously, \citet{lmc_streams} report a very similar
structure (dubbed S1), i.e. a large group of LMC stars in a fixed
narrow distance range, on the opposite side of the LMC. The top left
panel of their Figure 18 clearly shows that S1 stars do not follow the
disc's distance gradient. While the distance modulus range of S1 is
very restricted, it spans a wide range of angles on the sky. This
mostly flat, two-dimensional structure resembles the distribution of
the RR Lyrae on the Cloud's side facing the SMC and is suggestive of a
disc origin.

\subsection{The {\it Galex} litmus test and the Young Main Sequence bridge}

So far, for the RR Lyrae selection we have relied solely on the
photometry provided as part of the GSDR1. While measures of all sorts
have been taken to guard against contamination, at the moment it is
impossible to gauge with certainty the amount of spurious variability
supplied by the cross-match failures. However, there exists an
additional test which can help us to establish whether the discovered
bridge is genuinely composed of pulsating horizontal branch stars. RR
Lyrae are hot helium burning stars and as such occupy a narrow range
of broad-band color. Unfortunately, no deep optical survey provides a
wide-area coverage of the entire Magellanic system. Nonetheless, it
turns out that the brightest of the Magellanic RR Lyrae are seen by
the {\it Galex} space telescope.

Figure~\ref{fig:gaga_selection} gives the {\it Galex} DR7
\citep[GR7,][]{gr7} coverage of the $70^{\circ}\times70^{\circ}$
region around $(\LMS, \BMS)=(0^{\circ}, 0^{\circ})$. As can be seen in
the leftmost panel of the Figure, the {\it Galex} view of the Clouds
is very patchy. However, as shown in the second panel of the Figure,
most of the pixels around the LMC and the SMC have non-zero
completeness. The third panel of the Figure shows the Hess diagram
(density of sources in color-magnitude space) for the LMC sources
measured by both Gaia and the {\it Galex} AIS (the GaGa sample). As is
clear from the distribution of the previously identified RR Lyrae
stars (blue), the brightest of these are indeed present in GaGa and,
as expected, occupy a narrow range of $NUV-G$ color. The rightmost
panel of the Figure displays the familiar variability-magnitude
diagram for the GaGa stars within $10^{\circ}$ radius from the
LMC. Within the designated RR Lyrae box, an overdensity of objects is
visible. These stars are not only identified as variable by Gaia, but
also possess the $NUV-G$ color consistent with that of the RR
Lyrae. The latter is true even though no color cuts were applied to
select stars included in the diagram. This is because at the
magnitudes as faint as $G>18.5$ the {\it Galex} selection effects are
strong, and only stars with noticeable UV flux would be detected by
{\it Galex} (as seen in the third panel of the Figure). Nonetheless,
the selection of likely GaGa RR Lyrae candidates can be tightened if a
color cut - shown as the blue box in the third panel of the Figure -
is applied.

The left panel of Figure~\ref{fig:gaga_bridge} shows the distribution
of the GaGa RR Lyrae candidates (blue points) in the MS coordinate
system. Also shown are the contours of the GSDR1 RR Lyrae density
(black) corresponding to the selection shown in the left and center
panels of Figure~\ref{fig:bridge}. The completeness of the GaGa RR
Lyrae sample is truly minute, but its purity - thanks to the
additional color cut - is likely very high. The central part of the
LMC is missing from the GR7, and hence there is a large hole in the
distribution of blue points. At large angular distances from the LMC,
two prominent extensions of the GaGa RR Lyrae are traceable. The first
one is directly to the North from the LMC at
$10^{\circ}<\BMS<20^{\circ}$. This Northern RR Lyrae plume overlaps
with at least two recently discovered LMC sub-structures. First, a
section of the LMC's disc appears to be pulled in the direction of
increasing $\BMS$ as reported by \citet{Mackey2016}. Additionally, a
large tail of BHBs has been detected by \citet{lmc_streams},
stretching as far as $\BMS\sim25^{\circ}$ (S1 stream, e.g. their
Figure 6). The BHB density contours corresponding to the edge of the
LMC disc and the S1 structure are shown in purple. The second plume of
GaGa RR Lyrae is coincident with the GSDR1 bridge presented earlier
and reaches from the LMC to the SMC. Note that the SMC itself is not
very prominent, due to the drop in the GR7 AIS completeness at faint
magnitudes.

\begin{figure}
  \centering
  \includegraphics[width=0.45\textwidth]{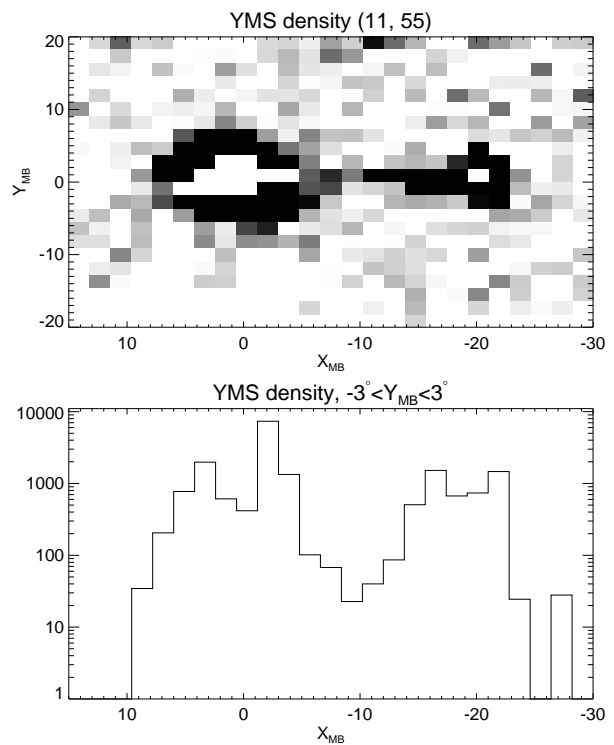}
  \caption[]{\small {\it Top:} Density of the GaGa YMS candidates in
    the Magellanic Bridge (MB) coordinate system in which both the LMC
    and the SMC lie on the equator. The pixel size is 1.8 degrees on a
    side. {\it Bottom:} Density of the GaGa YMS candidate stars along
    the MB equator. A foreground/background model (linear along MB
    latitude $Y_{MB}$ and different for each $X_{MB}$) is subtracted.}
   \label{fig:aligned_galex}
\end{figure}

The Hess diagram shown in the third panel of
Figure~\ref{fig:gaga_selection} also reveals a well populated Young
Main Sequence (YMS), seen as a cloud of stars with $NUV-G<2$ and
$G<19$. Taking advantage of this strong CMD feature and of the GaGa
wide coverage of the Clouds, we select YMS candidates using the CMD
box shown in black (without any cuts related to the stellar
variability as seen by \Gaia). The center panel of
Figure~\ref{fig:gaga_bridge} displays the density map of the GaGa YMS
candidate stars. Once again, the central parts of the LMC and the SMC
are missing due to the GR7 footprint irregularities. However, the
outer portions of the discs of both Clouds can be seen rather
clearly. Moreover, a narrow tongue of YMS stars appears to stick out
of the SMC and reach some $10^{\circ}$ across to the LMC. As the right
panel of the Figure clearly demonstrates, the RR Lyrae and the YMS
bridges are not coincident and follow distinct paths between the
Clouds.

Figure~\ref{fig:aligned_galex} presents the view of the YMS bridge in
the MB coordinate system. The top panel of the Figure shows that the
YMS bridge is a very narrow structure, which is nearly perfectly
aligned with the MB equator. We use a model identical to that
described in Section~\ref{sec:rrl_bridge} to extract the centroids and
the widths of the YMS bridge as a function of $\LMB$ and report these
in Table~\ref{tab:yms}. The bottom panel of the Figure shows the
density profile of the YMS bridge with background/foreground
contribution subtracted. The density along the bridge drops somewhat
in the periphery of the LMC, but otherwise is moderately flat with the
exception of a large excess of YMS stars in the Wing, i.e. on the side
of the SMC facing the LMC, at $-20^{\circ}<\LMB<-15^{\circ}$.

\begin{figure}
  \centering
  \includegraphics[width=0.48\textwidth]{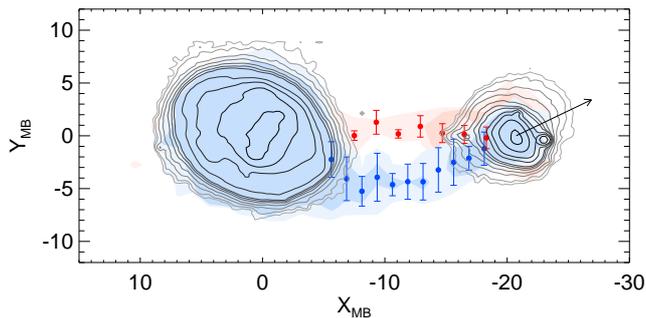}
  \caption[]{\small Comparison between the RR Lyrae (blue) and the YMS
    (red) bridges. The filled circles (error-bars) mark the centroids
    (widths) of each structure as extracted by the model (also see
    Tables~\ref{tab:rrl} and ~\ref{tab:yms}). The two structures are
    clearly offset on the sky at most $\LMB$ longitudes between the
    Clouds. However, they appear to connect to the SMC at
    approximately the same location on the east side of the dwarf. The
    contours give the all-star density distribution. The black arrow
    indicates the relative proper motion of the SMC with respect to
    the LMC. The YMS bridge connects to the Wing, while the RR Lyrae
    bridge connects to the southern portion of the S-shape. The
    conclusion is therefore inescapable that at least the portion of
    the RR Lyrae bridge closest to the SMC represents the dwarf's
    trailing tidal tail.}
   \label{fig:2bridges}
\end{figure}

Figure~\ref{fig:2bridges} compares the behavior of the RR Lyrae and
the YMS bridges as a function of the position on the sky. Throughout
most of the LMC-SMC span, the two bridges are clearly offset from each
other, with the largest angular separation being of order of
$\sim5^{\circ}$. At the distance of the bridge, this angular
separation corresponds to $\sim5$ kpc. Importantly, both connect to
the SMC at approximately the same location on the eastern side of the
dwarf. Note also the striking match between the all-star count
distribution (shown as contours) and the YMS/RR Lyrae bridge
density. This Figure demonstrates rather clearly that the RR Lyrae
bridge is the continuation of the lower part of the S-shape discussed
in Section~\ref{sec:starcounts}. We therefore conclude that the
portion of the RR Lyrae bridge closest to the SMC is the extension of
the dwarf's trailing arm. Given the line-of-sight distribution
discussed in Section~\ref{sec:rrl_bridge}, at $\LMB>-15^{\circ}$, the
bridge may be dominated by the LMC's stars. However we can not rule
out that some of the SMC's tidal debris reaches as far as the Large
Cloud or beyond.

\subsection{The HI bridge}
\label{sec:hi}

\begin{figure}
  \centering
  \includegraphics[width=0.48\textwidth]{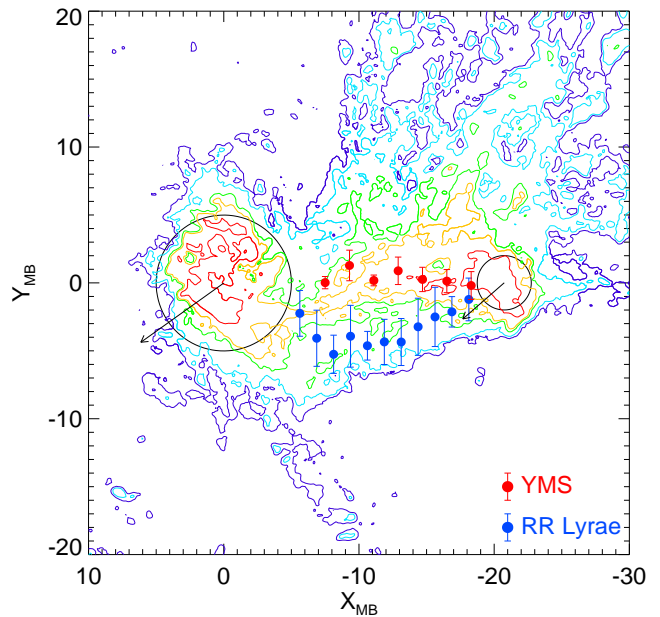}
  \caption[]{\small The three Magellanic bridges in the MB coordinate
    system. Contours give the density of the HI gas in the velocity
    range $100 < V_{\rm LOS} ($km s$^{-1} < 300$. Red, Yellow, Green,
    Blue and Purple contours correspond to gas column density of
    $(12.02, 3.18, 1.80, 0.46, 0.16) \times 10^{-20}$ cm$^{-2}$. Blue
    (red) filled circles with error-bars give the evolution of the
    centroid and the width of the RR Lyrae (YMS) bridge as a function
    of the MB coordinates. Blue filled circles with error-bars show
    the evolution of the centroid and the width of the RR Lyrae
    bridge. The LMC and the SMC are shown as large circles. Arrows
    give the proper motion vectors of the Clouds from
    \citet{nitya2013} in the MB coordinate system.}
   \label{fig:3bridges}
\end{figure}

Figure~\ref{fig:3bridges} shows the density map of neutral hydrogen in
and around the Clouds, based on the data from Galactic All-Sky Survey
\citep[GASS,
  see][]{gass_survey}\footnote{\url{https://www.astro.uni-bonn.de/hisurvey/gass/}}. This
represents the column density of HI gas with heliocentric velocities
$100$ km s$^{-1} < V < 300$ km s$^{-1}$ - a range that encompasses the
bulk of HI in the Magellanic system. As the Figure illustrates, the
regions of the highest gas density are coincident with the LMC and the
SMC (red contours). There is plenty of gas in between the Clouds as
well as trailing behind them (the Magellanic Stream), albeit at lower
density. Besides the Clouds themselves, the highest concentration of
HI appears to be in a narrow ridge-line structure, connecting the SMC
and the LMC, known as the Magellanic Bridge (mostly yellow contours).

It is obvious from the Figure that the GSDR1 RR Lyrae bridge is not
coincident with the main HI ridge of the inter-Cloud HI
reservoir. Instead, it is offset South-East, or, in other words, is
leading the gaseous bridge. Curiously, the Southern edge of the HI
distribution matches tightly the edge of the RR Lyrae bridge. The YMS
bridge, on the other hand, appears to sit almost exactly on the spur
of the HI from the SMC. The obvious conclusion from the distribution
of the young and the old stars in comparison to the neutral hydrogen
is that the YMS stars have formed in the gaseous bridge which was
stripped together with the RR Lyrae, but was pushed back (with respect
to the Clouds proper motion) by the ram pressure exerted by the
gaseous halo of the Galaxy. Also shown here are the arrows
corresponding to the proper motion vectors of the Clouds as measured
by \citet{nitya2013}. In Section~\ref{sec:ram} we will use the offset
between the young and old stars to estimate the gas density of the
Milky Way halo.

\section{Discussion and Conclusions}
\label{sec:disc}

\begin{figure*}
  \centering
  \includegraphics[width=0.98\textwidth]{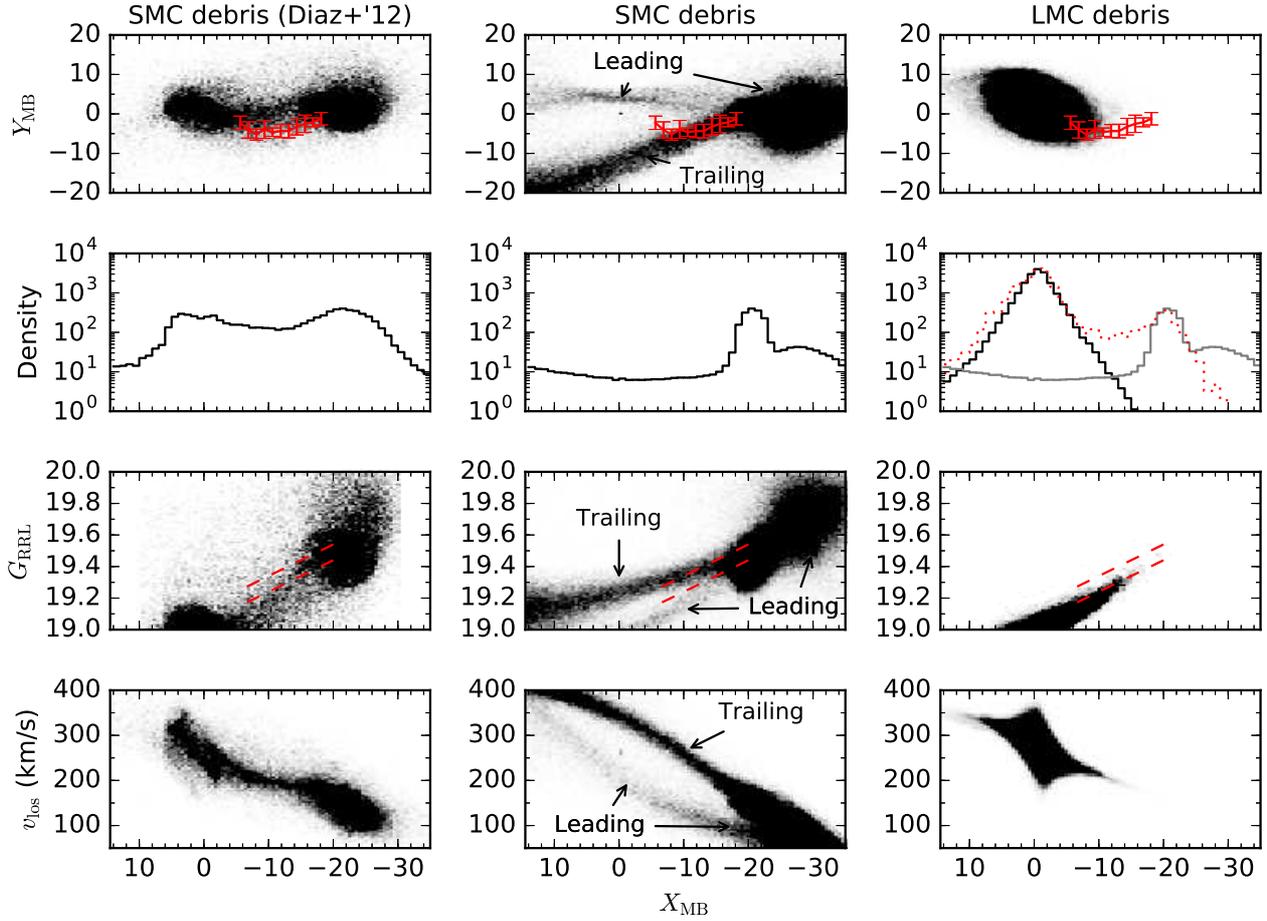}
  \caption[]{\small Debris from simulations of the SMC/LMC infall. The
    first column shows the debris from an SMC disruption similar to
    that in \protect\cite{Diaz2012}, the second column shows the SMC
    debris from a large number of simulations required to match the
    position of the bridge on the sky, the third column shows the LMC
    debris from \protect\cite{Mackey2016}. The rows show different
    observables of the debris: debris on the sky in Magellanic bridge
    coordinates, density of the debris along $X_{\rm MB}$, the G band
    magnitude of the debris, and the line of sight velocity of the
    debris. In the top row, we show the position of the old stellar
    bridge as a red line with error bars. In the third row, we show
    the observed distance gradient along the bridge with dashed red
    lines. Since the setup in \protect\cite{Diaz2012} was designed to
    match the HI bridge, the stars are above the old stellar bridge on
    the sky. The SMC debris shown in the middle column was required to
    match the old stellar bridge and as a result is a better fit. The
    debris has broadly the same distance gradient as observed although
    there is a large spread. The LMC debris in the right column shows
    that the tidally disrupting LMC disk can also provide a
    contribution in the region of the old stellar bridge. Note for the
    density in the LMC setup, we also show the density for the SMC
    setup (grey histogram), as well as the observed density from
    Fig. \protect\ref{fig:aligned} (red dotted histogram). Given that
    the simulated LMC density does not show any flattening, the
    observed flattening may be due to the SMC debris.}
   \label{fig:sim_debris}
\end{figure*}

\subsection{Comparison to the simulations}
\label{sec:sims}

In this Section we look to numerical simulations of the LMC and SMC
interaction in an attempt to interpret the RR Lyrae candidate
distribution presented above. In particular, we seek to find answers
to the following questions. Does the orientation of the RR Lyrae
bridge on the sky agree with the recently measured proper motions of
the Clouds?  What could be responsible for an inflection of the SMC's
trailing tail at $\LMB\sim-10^{\circ}$? What is the relative
contribution of each Cloud to the bridge density?  Here, we consider
three separate simulation setups: two which model the debris from the
SMC in the presence of the LMC, and one which only follows the LMC on
its orbit around the Milky Way. To produce realistically looking SMC
debris as it disrupts in the presence of the LMC, we use the modified
Lagrange cloud stripping technique of \cite{Gibbons2014} .

For the first simulation, we follow the setup of \cite{Diaz2012} with
an LMC represented by a Plummer sphere with a mass of $10^{10}
M_\odot$ and a scale radius of 3 kpc while the SMC is modelled as a
$3\times 10^9 M_\odot$ Plummer sphere with a scale radius of 2
kpc. The Milky Way is modelled using a three component potential made
up of a Miyamoto-Nagai disk, a Hernquist sphere bulge, and an NFW halo
\citep[see][for more details]{Diaz2012}. The SMC and LMC are rewound
from the final positions given in \cite{Diaz2012} for 3.37 Gyr and
then evolved to the present day. Material is stripped from the SMC
during its pericenters around the LMC with a rate given by a gaussian
with a dispersion of 50 Myr. This debris is shown in the left column
of Figure \ref{fig:sim_debris} where the rows show the debris on the
sky in MB coordinates, the density of the debris along $X_{\rm MB}$,
the G band magnitude of the debris, and the line of sight velocity of
the debris. \citet{Diaz2012} identified this particular combination of
parameters as it reproduced best the HI features of the Magellanic
Stream and Magellanic Bridge. As a result, it is not surprising that
the debris goes straight from the SMC to the LMC, unlike the bridge
seen in RR Lyrae (top left panel of Fig. \ref{fig:sim_debris}). The
distance gradient of this debris also seems somewhat off with respect
to what is observed since it quickly reaches a similar distance as the
LMC. This is because the debris from the SMC is accreted onto the LMC
in this setup. While this simulation provides only an approximate
match to the RR Lyrae bridge, it shows that it is possible for SMC
debris to attach onto the LMC. Thus, it is in principle possible that
a different setup could provide a better match to the RR Lyrae
observation presented here while also connecting to the LMC and hence
following the upturn in the bridge seen near the LMC. Note that there
exists important - albeit circumstantial - evidence as to the
existence of the SMC stellar debris inside the LMC \citep[see
  e.g.][]{Olsen2011}, which would superficially support the idea that
the RR Lyrae bridge extends uninterrupted all the way from the Small
to the Large Cloud.

In the second simulation where we study the SMC debris, we use a much
more massive LMC modelled as a Hernquist sphere with a mass of
$2.5\times 10^{11} M_\odot$ and a scale radius of 25 kpc. This heavy
LMC is in better agreement with the results of
e.g. \cite{Besla2010,jorge_lmc} as well as the constraint on the mass
enclosed within 8.7 kpc from \cite{vdm2014}. For the SMC we use a $2
\times 10^8 M_\odot$ Plummer sphere with a scale radius of 1 kpc. The
Milky Way is modelled as a 3 component potential
\texttt{MWPotential2014} from \cite{Bovy2015}. Using the updated
proper motion measurements for the LMC and SMC from \cite{nitya2013}, the line of sight velocities from \cite{vdm2002} and
\cite{Harris2006} for the LMC and SMC respectively, and the distances from \cite{LMCdist} and \cite{SMCdist} respectively, we sample the
position and velocity of the LMC and SMC. For each sampling, we rewind
the LMC and SMC for 3 Gyr, and then simulate the disruption of the
SMC. For each disruption, we construct a $\chi^2$ based on location of
the bridge on the sky and the bridge in distance and choose only the
simulations with $\chi^2/{\rm d.o.f.} < 1$. From 1000 simulations, we
find only 45 which satisfy the criteria suggesting that the location
of the bridge can be used to place tighter constraints on the proper
motion of the LMC and SMC. The combination of debris from these 45
simulations is shown in the middle column of Figure
\ref{fig:sim_debris}. This debris roughly matches the bridge's shape
on the sky although it does not display the turn-up seen in the data
near the LMC. Instead, it streams past the LMC to the South
East. While the trailing tail of the debris roughly matches the old
stellar bridge, the leading tail of the SMC reaches apocenter with
respect to the LMC and then heads back towards the LMC. Note that the
leading and trailing tails have different positions on the sky (the
trailing tail is below), different distances (the trailing tail is
farther away), and different line of sight velocities (the trailing
tail has a higher velocity). Also note that most of the stars in the
leading tail of the SMC are to the West of the SMC and more distant,
beyond the range RR Lyrae can be detected with GDR1. Furthermore, the
segment of the leading tail which appears as a stream has very few
stars compared to the trailing tail and thus may be too sparse to
detect with GDR1. Deeper future surveys, including GDR2, should be
able to detect the leading tail of the SMC. While beyond the scope of
this work, we note that the precise track of the trailing and leading
tail depend on the MW potential. Thus, future modelling efforts may be
able to use the old stellar bridge to get a constraint on the MW halo.

We note that these simulations of the SMC debris neglect several
important effects. First, we do not account for the dynamical friction
of the SMC in the presence of the LMC. If dynamical friction were
included, the SMC would have been farther away in the past and would
have stripped less. As a result, the length of the streams in
Figure~\ref{fig:sim_debris} can be reduced depending how effective
dynamical friction is. Second, the Lagrange cloud stripping technique
was not designed to correctly model the density along the stream with
respect to the dwarf, rather it is designed to match the stream track
in position and velocity. Thus, the peaky SMC density in the middle
column of Figure~\ref{fig:sim_debris} should not be
over-interpreted. As a test of the second set of simulations,
especially given the small pericenters of the SMC with respect to the
LMC, we have run several N-body simulations with \textsc{gadget-3}
\citep[similar to \textsc{gadget-2}][]{Springel2005}. In these
simulations, the LMC is modelled as a particle sourcing a Hernquist
potential and the SMC is modelled as a live Plummer sphere with $10^5$
particles. The pattern of debris looks almost identical showing that
the Lagrange cloud stripping technique works well.

Finally, we have a simulation of the evolution of the LMC disk under
the Galactic tides, identical to that in \cite{Mackey2016}. Unlike the
previous two setups, this simulation contains no SMC and thus neglects
the perturbations that it can impart on the LMC
\citep[e.g.][]{Besla2012,Besla2016}. However, it does capture the
response of the LMC to the Milky Way. This setup involves a live two
component N-body LMC (disk+dark matter) disrupting in the presence of
a live three component Milky Way \citep[see][for more
  details]{Mackey2016}. The stars from the LMC disk are shown in the
rightmost column of Figure~\ref{fig:sim_debris}. The LMC disk debris
stretch out to the location of almost the entire bridge. In addition,
the distance gradient matches the bridge. Thus, it is likely that some
of the bridge, and perhaps the upturn near the LMC, is due to debris
from the LMC. This is emphasised in the 2nd row, 3rd column panel of
Figure \ref{fig:sim_debris} where we show the density of the LMC
(black histogram), the density of the SMC debris from the middle
column (grey histogram), and the observed density of RR Lyrae from
Figure \ref{fig:aligned} (black histogram from 2nd column). We see
that the observed density matches the LMC quite well for $X_{\rm MB} <
7^\circ$, after which it flattens out. The flattening beyond $X_{\rm
  MB} > 7^\circ$ is likely due to SMC material. Note that the
simulated density has been scaled to match the observed density peak
near the LMC and SMC.

\begin{figure*}
  \centering
  \includegraphics[width=0.98\textwidth]{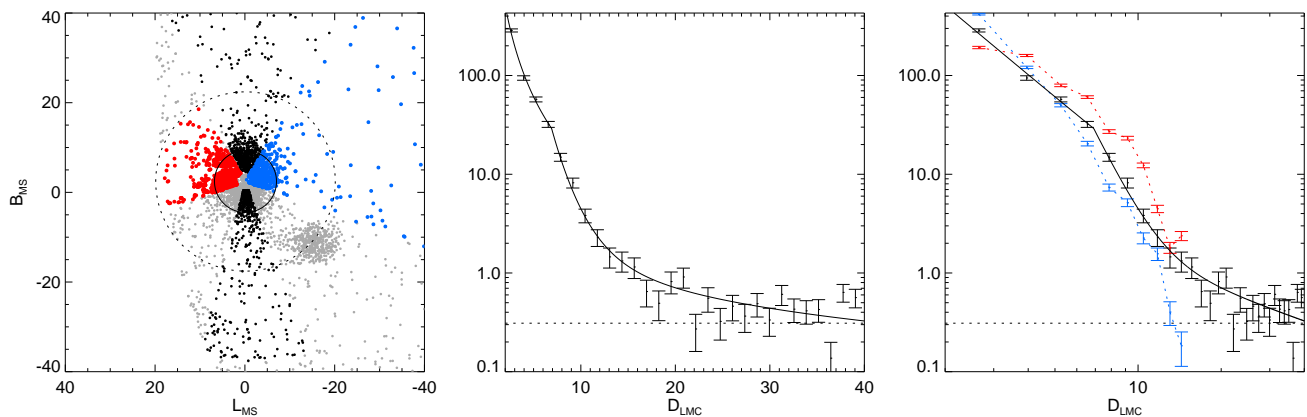}
  \caption[]{\small {\it Left:} Distribution of $\sim$6,000 Gaia DR1
    RR Lyrae candidates in the Magellanic Stream coordinate
    system. These were selected with ${\rm Amp}>-0.65$ and $\log({\rm
      AEN})<-0.2$ cuts. Sub-sets of stars colored black and blue are
    used for the radial profile modeling reported in the center and
    right panels. The solid circle marks the break radius of $\sim
    7^{\circ}$, while the dotted circle simply shows the $20^{\circ}$
    boundary around the LMC. {\it Center and Right:} LMC RR Lyrae
    radial density profiles (black data-points with error-bars) and
    the maximum-likelihood broken power-law model (black line). Red
    and blue data-points correspond to the leading and trailing parts
    of the LMC as shown in the Left panel of the Figure.}
   \label{fig:profile}
\end{figure*}

The simulations show that around the LMC, both the Large and the Small
Cloud can naturally produce debris which is closely aligned with the
RR Lyrae bridge on the sky and in distance (right two columns of
Fig. \ref{fig:sim_debris}). Fortunately, these debris have different line
of sight velocity signatures with the SMC debris having a much higher
velocity, $\sim 50$ km/s, at the same $X_{\rm MB}$. Thus,
spectroscopic follow-up of the stars in the RR Lyrae bridge should
allow us to test whether the debris is partly made up of SMC and LMC
debris and if there is a transition between the two. Note that the
entire bridge could also come from SMC debris which would require an
upturn near the LMC. Although the models shown in the middle column of
Figure \ref{fig:aligned} do not show this behavior, a larger search of
the parameter space may uncover SMC debris somewhere between the first
and second column. The radial velocity signature of this debris would
presumably connect smoothly from the SMC to the LMC and not exhibit
two distinct populations.

Based on the analysis of the simulations presented above, we conclude
that at least at $\LMB<-10^{\circ}$, the SMC trailing tail contributes
most of the material to the RR Lyrae bridge. Additionally, as
explained in \citet{Diaz2012} and shown in Figure~\ref{fig:sim_debris},
there exists a counterpart to the
trailing arm: the SMC's leading arm, mostly on the opposite side of
the Cloud, albeit it is not arranged as neatly as the
trailing. Instead, it is bending away from the observer and around the
SMC, thus appearing much shorter on the sky as well as extending
further along the line of sight. While a segment of the leading tail
looks stream-like, most of the stars in the leading tail are in the
field of debris to the West of the SMC. Note that all simulations
discussed so far predict some of the SMC tidal debris outside of the
main area of the RR Lyrae bridge. Much of the stripped material
appears to lead the LMC. How far it can be flung out is likely
controlled by the size of the SMC's orbit.

\subsection{The stellar outskirts of the LMC}
\label{sec:lmc_halo}

The focus of this paper is on the tidal tails of the SMC, in
particular the trailing arm, which - when traced with RR Lyrae - has
the appearance of a bridge connecting the Small Cloud to the Large. In
this Subsection, we concentrate on the properties of the distribution
of the RR Lyrae residing in and around the LMC.

The left panel of Figure~\ref{fig:profile} shows the locations (in the
MS coordinate frame) of individual RR Lyrae candidate stars selected
using the strict version of the cuts presented in
Equation~\ref{eqn:selection}, both in $\Amp$ and $\AEN$. This sample
is then divided into four groups based on the star's azimuthal angle
(indicated with color). We use the stars in the blue and black groups
to model the LMC's radial density profile, but avoid the red group as
it runs into the regions of low Galactic latitude as well as the grey
group as it contains the SMC and its trailing tail (i.e. the
bridge). The resulting radial density profile is shown in linear
(logarithmic) scale in the middle (right) panel of the Figure. Both
panels demonstrate a clear change in the behavior of the stellar
density between $5^{\circ}$ and $10^{\circ}$ from the LMC's center
where the star count rate drops noticeably. Also note that the RR
Lyrae distribution extends as far as $20^{\circ}$ from the center of
the LMC, if nor further.

Motivated by the behavior of the LMC stellar density, we model the
distribution of the candidate RR Lyrae stars with a broken power-law
(BPL) (best-fit solution shown as solid black curve). In a BPL model,
the density distribution is described with a simple power law, but the
power-law index is allowed to change at the break radius. The two
power law indices (inner and outer), the break radius and the (flat)
background contribution are the free parameters of this model. Note
that similar BPL models have been used successfully to describe the
density profile of the Milky Way stellar halo \citep[see
  e.g.][]{Sesar2011,broken,Xue2015}. The maximum-likelihood model of
the 1D angular distribution of the RR Lyrae candidate stars in the
black and blue groups places the break at the radius of
$6.91^{\circ}\pm 0.34^{\circ}$. The inner power law index is $2.36 \pm
0.08$, while the outer power law index is $5.8 \pm
0.6$. \citet{breaks} put forward a simple explanation of radial
density breaks in the stellar halos around Milky Way-like galaxies. In
their picture, the breaks emerge if the stellar halo is dominated by a small number of massive progenitors that are accreted at reasonably early times.

Before we speculate as to the origin of the LMC's stellar halo, it is
prudent to point out some of the drawbacks of the above modeling
exercise, most notably, the assumption of spherical symmetry and the
alignment of the disc and the halo. For example, if the center of the
LMC's disc is offset from the center of the LMC's halo, the
(mis-centered) radial density profile will acquire an artificial
``scale''. Furthermore, if beyond a certain radius the LMC's RR Lyrae
distribution is flattened, the resulting number count profile may look
``broken''. There is some modest evidence for an elongation of the LMC
as traced by the RR Lyrae as can be seen from the comparison of the
black and blue lines in the right panel of
Figure~\ref{fig:profile}. The black line gives the count for both
black and blue points (thus indicating the average behavior) in the
left panel, but the blue line shows the properties of the blue points
only. The blue profile is systematically above the black at small
radii and sits below it at large angular distances. This may be
because - in the MS system - the LMC is stretched vertically (or
squashed horizontally).

Apart from the hints of a possible flattening, the RR Lyrae
distribution also shows signs of asymmetry. This can be gleaned from
the shape of the red line in Figure~\ref{fig:profile} as compared to
the overall profile (given in black). There appears to be a strong
excess of RR Lyrae on the leading (with respect to its proper motion)
side of the Cloud. This discovery agrees well with the most recent map
of the Magellanic Mira stars presented in \citet{alis_mira} and
discussed in more detail in Section~\ref{sec:mira}.

The preliminary (due to the contaminated and largely incomplete RR
Lyrae sample considered here) results can be compared to some of the
recent attempts to measure the radial density profile of the LMC. The
NOAO's Outer Limits Survey (OLS) obtained a large number of deep
images of the LMC, in which the dwarf's Main Sequence population can
be traced as far as $\sim16^{\circ}$ from its center
\citep[see][]{Saha2010}. The MS counts in the OLS sample follow an
exponential profile. However, the spectroscopically confirmed red
giant branch (RGB) stars from the survey of \citet{Majewski2009}
appear to have a break at the distance of $\sim9^{\circ}$ in the
radial density profile. The RGBs consistent with the LMC population
can be traced as far as $\sim23^{\circ}$ in agreement with the RR
Lyrae distribution discussed above, albeit beyond the break, instead
of steepening (as found here), their density profile flattens. Last
year, Dark Energy Survey (DES) provided a deep and continuous view of
a small portion of the LMC. The analysis of the DES data can be found
in \citet{Balbinot2015} and \citet{Mackey2016}. In agreement with
\citet{Saha2010}, \citet{Balbinot2015} find that the LMC stellar
content is dominated by disc population with a truncation radius of
$\sim13$ kpc. An independent examination of the DES data is reported
in \citet{Mackey2016} who detect i) pronounced East-West asymmetry in
the Cloud's radial density profile as well as ii) a strong evidence
for a very diffuse stellar component reaching beyond $\sim20^{\circ}$
from its center. Both of these findings appear to be in excellent
agreement with the results based on the GSDR1 RR Lyrae sample
presented here. It however remains unclear what morphological
component is responsible for the extended envelope of stars around the
Large Cloud, the disc or the halo; where the interface between the two
lies, and if the stellar halo exists what processes are responsible for
its creation.

\subsection{Comparison to Mira results}
\label{sec:mira}

\citet{alis_mira} present a large number of candidate Mira stars in
the vicinity of the Magellanic Clouds. This is a new sample of Mira
constructed using a combination of \Gaia, 2MASS and WISE colors as
well as the \Gaia\ variability statistic $\Amp$ and is estimated to
have very low levels of contamination. Note that the sample of stellar
tracers discussed in \citet{alis_mira} includes both Mira and
Semi-Regular Variables shown in the second and the third panels of the
top row of Figure~\ref{fig:variables}. Around the LMC, GSDR1 Mira
stars are seen at reasonably large angular separations from the LMC,
most prominently in the North, where they overlap with the ``stream''
discovered by \citet{Mackey2016}, in the South where they overlap with
the beginning of the extension mapped by the RR Lyrae presented
here. However, there is no indication of the Mira presence in the area
covered by the RR Lyrae bridge, i.e. in between the Clouds. While this
might be a reflection of the stellar population gradients in the SMC
disc, this could actually be simply due to the very low stellar
density in the bridge. This latter explanation is perhaps preferred as
the Mira distribution in the SMC does show noticeable excess - see
Figure 8 of \citet{alis_mira} - on the ends of the S-shape structure
traced by the \Gaia's raw star counts, i.e. in the densest portions of
the two tidal tails. Additionally, there are several Mira candidates
in the East of the LMC (in the MS coordiante system), where they match
the RR Lyrae excess discussed in Section~\ref{sec:lmc_halo}.

The Mira stars in \citet{alis_mira} can also be traced to regions of
the sky away from the Magellanic Clouds (see their Figure 11). In
particular, some of the Mira identified \textit{above} the Galactic
plane at $l \sim -90^\circ$ could be associated with the SMC debris
leading the Clouds. Searches for other stellar populations (like RR
Lyrae stars) in the region of the predicted far-flung Magellanic
debris will help confirm this result, and will further test models of
the SMC/LMC infall.

\subsection{RR Lyrae bridge in  the OGLE IV observations}
\label{sec:ogle}

The OGLE IV's sample of the Magellanic RR Lyrae
\citep[see][]{ogle_rrl} is both more complete and more pure compared
to the one analysed here. The only advantage of the GSDR1 data is the
unrestricted view of the both Clouds and the area between and around
them. On inspection of the top panel of Figure 1 of \citet{ogle_rrl},
it is evident that i) OGLE IV has detected the RR Lyrae in the
trailing arm of the SMC and ii) it is impossible to interpret it as a
narrow bridge-like structure using the OGLE data alone as it lies at
the edge of the survey's footprint.

\begin{figure}
  \centering
  \includegraphics[width=0.48\textwidth]{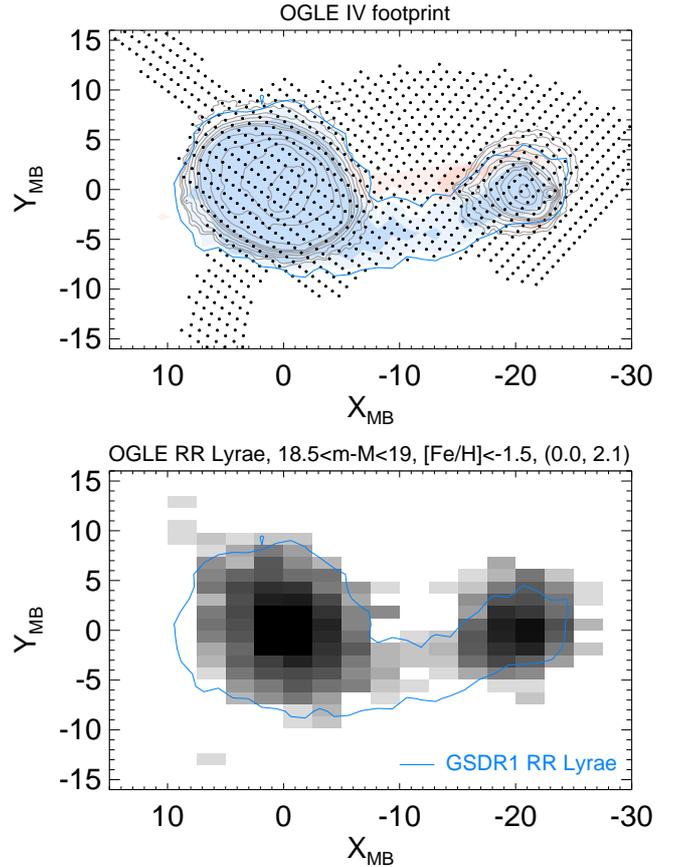}
  \caption[]{\small {\it Top:} OGLE IV footprint in MB
    coordinates. Locations of individual survey fields are marked with
    small black dots. Underlying are the density contours discussed
    earlier in the paper (see Figure~\ref{fig:2bridges}).  {\it
      Bottom:} Logarithm of the density of the OGLE IV RR Lyrae with
    $18.5<m-M<19$ and [Fe/H]$<-1.5$. A narrow structure connecting the
    two Clouds is clearly visible, matching the location, the extent
    and the breadth of the GSDR1 RR Lyrae bridge.}
   \label{fig:ogle}
\end{figure}

Further evidence as to the existence of the old tidal debris in the
OGLE data can be found in \citet{skowron_bridge}. Their Figures 11 and
13 show the distribution of the top red-giant branch and the bottom
red-giant branch stars, corresponding to the intermediate and the old
populations respectively. While the intermediate-age stars (their
Figure 11) do not trace any striking coherent structure in the
inter-Cloud space, an uninterrupted bridge of old stars is obvious at
the edge of the footprint (their Figure 13). Once again,
unfortunately, the limited field of view does not allow an estimate of
the actual width of the structure.

Most recently, \citet{ogle_rrl2} presented a detailed study of the
structure of the two Clouds and the area between them using a
sub-sample of the OGLE IV RR Lyrae. After selecting only RRab
pulsators and culling objects with uncertain lightcurve shape
parameters from the original sample of $\sim$45,000 stars, the authors
end up with $\sim22,000$ RR Lyrae.  The results of this study can be
summarized as follows. The RR Lyrae density distributions of the LMC
and the SMC can be described with families of nested ellipsoids. In
the LMC, \citet{ogle_rrl2} detect a noticeable twist in the
orientation of the major axes of the ellipsoids as a function of the
distance away from the Cloud's center, while the density field of the SMC
appears much more regular and symmetric. Overall, no strong
irregularities or asymmetries have been reported for either of the
Clouds. With regards to the inter-Cloud space, the paper announces the
presence of a small number of RR Lyrae, but nothing similar to a
coherent structure discussed here.

At a first glance, some of the conclusions reached in
\citet{ogle_rrl2} appear inconsistent with the sub-structure
detections from the GSDR1 data. Around the LMC, this includes the
Northern structure, i.e. overlapping with the \citet{Mackey2016}
``stream'' and the S1 BHB/RR Lyrae stream \citep[][]{lmc_streams}, the
Eastern excess of RR Lyrae (see Section~\ref{sec:lmc_halo} of this
paper) as well as the Southern LMC extension, which could be
responsible for as much as a half of the bridge we see in GSDR1. None
of these entities seem to be confirmed with the OGLE IV data. However,
the explanation for this seeming disagreement might be rather simple:
all of the sub-structures mentioned above lie in the periphery of the
Cloud, and thus do not fall within the OGLE IV's footprint shown in
the top panel of Figure~\ref{fig:ogle}. This is certainly true for the
Northern and Eastern parts of the LMC. The OGLE IV coverage of the
Southern portion of the Cloud is broader, but even there, the
structures reported here sit right at the edge of the footprint.

To compare the properties of the GSDR1 and OGLE IV RR Lyrae more
directly, we build a map of the density distribution of a sub-sample
of RRab pulsators from \citet{ogle_rrl}. More precisely, we select RR
Lyrae with well-determined light-curve shapes, i.e. those with errors
on the $\phi_{31}$ and $\phi_{21}$ parameters smaller than
0.5. Additionally, we require the stars to lie at distances larger
than that of the LMC but smaller than that of the SMC,
i.e. $18.5<m-M<19$. Finally, we only plot metal-poor RR Lyrae, namely
those with [Fe/H]$<-1.5$. The number of RR Lyrae satisfying all of the
conditions above is approximately $\sim3,700$ (which is approximately
1/5 of all RR Lyrae of ab type with good lightcurves within the
designated distance range) and their density distribution is shown in
the bottom panel of Figure~\ref{fig:ogle}. According to
\citet{ogle_rrl2}, the OGLE IV RR Lyrae cover a broad range of
Magellanic Bridge latitudes $\BMB$. However, as clear from the Figure,
the metal-poor subsample traces exactly the same narrow structure
mapped out by the GSDR1 RR Lyrae candidates. We, therefore, conclude
that the two distributions are in agreement with each other, albeit
for a few pixels in the OGLE map at low $\BMB$ with depleted star
counts, which are likely due to the effects of the survey's
footprint.


\subsection{Density of the Milky Way's hot corona}
\label{sec:ram}

Above, we have shown that there are two bridges between the SMC and
LMC: a gaseous bridge which contains YMS stars and a bridge containing
old stars (e.g. Fig~\ref{fig:3bridges}). The gaseous bridge trails the
old stellar bridge relative to the direction in which the LMC/SMC are
moving by $\sim 5$ kpc. Since both bridges connect to the SMC at the
same location, it is likely that both bridges come from material
stripped from the SMC during the same previous pericenter about the
LMC. Interestingly, the relative proper motion of the SMC with respect
to the LMC is aligned with the stellar bridge suggesting the bridge is
the trailing arm (e.g. Fig~\ref{fig:2bridges}). In
Section~\ref{sec:rrl_bridge}, we hypothesised that the offset between
the bridges is likely caused by the additional ram pressure which is
being exerted on the gaseous bridge by hot gas in the Milky Way
halo (corona). Equipped with the offset in the bridges, $\Delta x$, the time
since material was stripped, $\Delta t$, the relative velocity of the
LMC and the MW, $v_{\rm rel}$, and the column density of neutral gas in the
bridge, $N_{\rm MB}$, it is possible to roughly estimate the gas
density of the hot corona of the Milky Way, $\rho_{\rm cor}$.

The ram pressure on the gaseous bridge is given by $\rho_{\rm cor}
v_{\rm rel}^2$. If we consider a block of the gaseous bridge with area
$dA$ facing the oncoming gas and length $dl$, the force on this block
from ram pressure is $\rho_{\rm cor} v_{\rm rel}^2 dA$ and the mass of
the block is $dM =\rho_{\rm MB} dA dl$. If we further assume that the
extent of the gaseous stream perpendicular to its track is roughly
similar in both directions, which is justified if it is a stream, then
the column density and density of the bridge are related by $N_{\rm
  MB} \sim n_{\rm MB} dl$, where $n_{\rm MB}$ is the average number
density of hydrogen atoms in the bridge.  As a consequence, the mass
of the gas block is $N_{\rm MB} \mu_{\rm MB} m_{\rm p}dA$, where
$\mu_{\rm MB}=1.33$ is the atomic weight assuming that the gas in the
bridge is neutral and that the gas is made up of the universal fractions of hydrogen and helium, and $m_{\rm p}$ is the proton
mass. This gives an acceleration of
\begin{equation}
a \sim \frac{n_{\rm cor} \mu_{\rm cor} v_{\rm rel}^2}{N_{\rm MB} \mu_{\rm MB} }
\end{equation}
where we have written the coronal density in terms of the number density as $\rho_{\rm
  cor}=n_{\rm cor}\mu_{\rm cor}m_{\rm p}$ with an atomic weight
$\mu_{\rm cor}\simeq 0.6$ since this medium is hot and largely ionised
\citep{Miller&Bregman2015}.  Assuming that the gaseous and old stellar
bridge have been exposed to the ram pressure for some time $\Delta t$,
at the present they will have an offset of
\begin{equation}
\Delta x \sim \frac{n_{\rm cor} \mu_{\rm cor} v_{\rm rel}^2}{2 N_{\rm MB} \mu_{\rm MB}} \Delta t^2 .
\end{equation}
Solving for the coronal number density we find
\begin{equation}
n_{\rm cor}  \sim \frac{2 \mu_{\rm MB} N_{\rm MB} \Delta x}{\mu_{\rm cor} v_{\rm rel}^2 \Delta t^2} .
\end{equation}
Plugging  in numbers  of $v_{\rm  rel} \sim  350$ km/s  (based on  the
observed proper motion and radial velocity of the LMC), $\Delta x \sim
5$ kpc (from  the measured offset), $\Delta t \sim  200$ Myr (from the
typical time the  simulated LMC/SMC enter the region within  60 kpc of
the MW), and  $N_{\rm MB} \sim 2\times10^{20} {\rm  cm}^{-2}$ from the
observed HI column density of the dense part of the bridge as shown in
Figure~\ref{fig:3bridges} \citep[see also][]{Putman2003}, we find
\begin{equation} 
n_{\rm cor} \sim 3\times 10^{-4} {\rm cm}^{-3} 
\end{equation}
This rough estimate is consistent with previous estimates based on ram
pressure effects on Milky Way satellites: $1.3-3.6\times 10^{-4} {\rm
  cm}^{-3}$ \citep{gatto_et_al_2013} and $0.1-10\times10^{-4} {\rm
  cm^{-3}}$ \citep{grcevich_putman_2009}, as well as estimates based
on the distortion of the LMC disc: $0.7-1.5\times10^{-4} {\rm
  cm}^{-3}$ \citep{salem_et_al_2015}.
Finally, it satisfies the upper limit for the average electron
number density between us and LMC, $\langle n_{\rm e}\rangle\simeq 5\times
10^{-4}{\rm cm}^{-3}$, determined using dispersion measures from pulsars 
on the LMC \citep{Anderson&Bregman2010}.

Note that this estimate comes with several additional caveats. First,
we have assumed that both the old stellar bridge and the gaseous
bridge are the trailing tail of the SMC debris while they could, in
principle, represent leading and trailing arms of the stream or even
different wraps. However, given Figure~\ref{fig:2bridges} which shows
that the relative proper motion of the SMC with respect to the LMC is
aligned with the old bridge suggesting it is the trailing tail, and
the results of \cite{Besla2012,Diaz2012} which both find that the HI
bridge is well modelled by the trailing tail of SMC debris, we think
this is a reasonable assumption. Second, we have assumed that the HI
gas bridge and the old stars are stripped from the SMC with the same
velocity. However, the gas in the SMC will feel additional ram
pressure from the gas in the LMC so the two bridges may look different
even before accounting for ram pressure from the Milky Way gas. Finally,
we have assumed that the ram pressure simply displaces the gas relative to the stars.
In reality, the high relative velocities will give rise to Kelvin-Helmholtz instabilities and turbulence which
will compress and shred the gas, making the gaseous bridge wider and more diffuse. An in-depth understanding of 
how the HI bridge interacts with the ambient material would require a full 
hydrodynamical treatment of the system, which is beyond the scope of this paper.

We stress that while this is an extremely simple estimate, it shows
that the offset is a powerful probe of the gas density in the Milky
Way halo. In the future, realistic hydrodynamic simulations of an
LMC/SMC pair accreted onto the Milky Way which address the caveats
above should be able to provide much more precise estimates.

\subsection{Conclusions}

We have used the \Gaia\ DR1 photometry catalog {\texttt GaiaSource} to
study the outer environs of the Small and Large Magellanic Clouds. As
part of our investigation, we demonstrate that genuine variable stars
can be detected across the whole sky relying only on the \Gaia's mean
flux and its associated error. In this work, we concentrate on the
sample of candidate Magellanic RR Lyrae identified using GSDR1 data
alone. Unsurprisingly, giving the limited information in hand, the
sample's completeness and purity are low compared to the datasets
where lightcurve and/or color information is available. The major
stumbling block unearthed as part of our analysis is the spurious
variability caused by the (likely) failures of the object cross-match
algorithm used for the GDR1 creation. Nevertheless, through a series
of tests, we demonstrate that the faint features we discover around
the Clouds are bona fide. The results of this work can be summarized
as follows.

\begin{itemize}

\item Even with the GDR1 {\texttt GaiaSource} star counts alone, the
  outer density contours of the SMC can be shown to twist noticeably,
  forming a familiar S-shape, symptomatic of tidal
  stripping. Furthermore, the twist is aligned with the relative
  proper motion of the SMC with respect to the LMC. Thus, we
  conjecture that the LMC is the likely cause of the disruption. Using
  the SMC's proper motion relative to its violent neighbor, we
  classify the tail pointing towards the Large Cloud as trailing and
  the one on the the opposite side of the Small Cloud as leading.

\item The distribution of the RR Lyrae reveals a long and narrow
  structure connecting the two Clouds. This RR Lyrae ``bridge'' joins
  the SMC exactly where the base of the trailing tail can be seen in
  the all-star density map described above. To verify the nature of
  the bridge, we use GaGa, a combination of \Gaia\ and {\it Galex}
  photometry. The purity of the GaGa RR Lyrae subset is much higher
  than that of the original GSDR1 sample thanks to the UV-optical
  color cut applied. There are only two prominent structures visible
  in the GaGa RR Lyrae distribution. The first one is the bridge
  between the Clouds and the second one is the counterpart of the
  Northern LMC's extension traced previously by \citet{Mackey2016} and
  \citet{lmc_streams}.

\item The GaGa photometry allows for an efficient selection of Young
  Main Sequence stars at the distance of the Clouds. Using the GaGa
  YMS sample, we build a high-resolution map of a narrow bridge
  composed of stars recently formed within the neutral hydrogen
  stripped from the SMC. In agreement with previous studies, e.g. most
  recently by \citet{skowron_bridge}, the YMS bridge shows nearly
  perfect alignment with the HI bridge. However, the RR Lyrae bridge
  is offset from both the YMS stars and the HI by some
  $\sim5^{\circ}$.

\item Assuming a constant absolute magnitude to the GSDR1 RR Lyrae, we
  study the 3D structure of the bridge. It appears that at many
  positions along the bridge, two structures at different
  line-of-sight distances can be discerned, one at the mean distance
  of the LMC and one with distances evolving smoothly from the SMC to
  the LMC. Taking into account the evolution of the selection
  efficiency with magnitude, we estimate that each structure
  contributes similar number of RR Lyrae around the mid-point of the
  bridge. Therefore, the RR Lyrae bridge is a composite structure,
  consisting of two stellar streams, one from the LMC and one from the
  SMC.

\item Simulations of the Magellanic in-fall appear to be in a broad
  agreement with the observations presented here. They also help to
  clarify some of the uncertainties in the interpretation of the
  \Gaia\ data. At $\LMB<-10^{\circ}$, the RR Lyrae bridge is mostly
  composed of the SMC stellar debris. This part of the bridge is
  simply the Cloud's trailing tail, while its leading tail is
  compressed on the sky and stretched along the line of sight. The
  simulations confirm that the LMC stars can contribute significantly
  to the inter-Cloud RR Lyrae density to cause an up-turn of the
  bridge towards the LMC at $\LMB>-10^{\circ}$. Thus, the above
  hypothesis that a significant part of the RR Lyrae bridge detected
  here is an extension of the LMC is reinforced.  Curiously, the
  obvious distance gradient in the LMC leaves the nature of the
  stellar structure stretching out of the dwarf at constant $G\sim19$
  to $\LMB=-15^{\circ}$ rather enigmatic.

\item Our results are consistent with the picture of the Clouds
  painted with Mira-like stars as presented in \citet{alis_mira}. For
  example, there is strong evidence that, similarly to GSDR1 RR Lyrae,
  the Mira stars trace the LMC as far as $\sim20^{\circ}$ from its
  center. Furthermore, an excess of Mira stars is detected in the North, the
  South and the East of the Large Cloud, thus matching the RR Lyrae
  sub-structures discussed above. Around the SMC, while no visible
  bridge connecting the Small Cloud to the Large is discernible, there
  appear to be groups of Mira accumulating at the ends of the S-shape
  structure.

\item Finally, using the offset between the RR Lyrae and the HI
  bridges, we provide a back-of-the-envelope estimate of the density
  of the hot gaseous corona of the Milky Way. Under the assumption
  that both neutral hydrogen and the stars were stripped from the SMC
  at the same time, the MW halo ought to have density of order of
  $\rho_{\rm MW} \sim 3\times 10^{-4} {\rm cm}^{-3}$ to provide the
  necessary ram pressure to push the HI gas $\sim5^{\circ}$ in the
  trailing direction. Our calculation is simple, but, importantly, is
  consistent with previous estimates. We believe, therefore, that if
  the discovery of the stellar tidal tails of the SMC is confirmed, an
  improved version of the ram-pressure argument presented here can be
  used to put tight constraints on the amount of hot gas within the
  viral volume of the Galaxy.
  
\end{itemize}

We envisage that in the nearest future, the true nature of the RR
Lyrae bridge uncovered here will be verified with the help of
follow-up observations. In fact, this can be done using the data from
the \Gaia\ satellite itself, i.e. that contained within the Data
Release 2, which will provide individual stellar colors as well as
robust stellar variability information. Bearing in mind the complex
interwoven 3D structure of the debris distribution between the Clouds,
it will undoubtedly be beneficial to obtain deep broad-band photometry
of the region. This should help to disentangle the individual
contributions of the LMC and the SMC. As illustrated above, different
numerical simulations of the Clouds' in-fall predict distinct patterns
in the line-of-sight velocity space. Therefore, the wide-area
spectroscopic survey of the Clouds' periphery will be an important
next step in deciphering the history of their interaction. Given the
unexpected richness of the GDR1, it is certain that the future
\Gaia\ releases are bound to be truly revolutionary, not only for the
inner Galaxy but for its outer fringes too.

\input{lmc_bridge_table_rrl}
\input{lmc_bridge_table_yms}

\section*{Acknowledgments}

The authors are indebted to the \Gaia\ team in general and in
particular to Giorgia Busso, Alcione Mora and Anthony Brown for the
swift and expertly support they have been providing. It is a pleasure
to thank Gurtina Besla and Justin Read for sharing their wisdom
regarding the simulations and observations of the Magellanic HI. We
also wish to thank Igor Soszy\'nski for the advice on the OGLE
variable star data this study has benefited from.

This project was developed in part at the 2016 NYC Gaia Sprint, hosted
by the Center for Computational Astrophysics at the Simons Foundation
in New York City and in part at the Dark Matter Distribution in the
Era of Gaia Workshop, hosted by NORDITA.

This work has made use of data from the European Space Agency (ESA)
mission {\it Gaia} (\url{http://www.cosmos.esa.int/gaia}), processed
by the {\it Gaia} Data Processing and Analysis Consortium (DPAC,
{\small
  \url{http://www.cosmos.esa.int/web/gaia/dpac/consortium}}). Funding
for the DPAC has been provided by national institutions, in particular
the institutions participating in the {\it Gaia} Multilateral
Agreement.

The authors thank the team of The Parkes Galactic All-Sky Survey for
making their data public.

The research leading to these results has received funding from the
European Research Council under the European Union's Seventh Framework
Programme (FP/2007-2013) / ERC Grant Agreement n. 308024. V.B.,
D.E. and S.K. acknowledge financial support from the ERC. S.K. 
also acknowledges the support from the STFC (grant ST/N004493/1).
A.D. is supported by a Royal Society University Research Fellowship.

\label{lastpage}

\end{document}

%% file: lmc_bridge_table_rrl.tex
\begin{table*}
\caption{Properties of the RR Lyrae bridge.}
\begin{tabular}{|c c c c c c c c c c c c|}
\hline
$X_{\rm MB}$& -5.6 &  -6.9 &  -8.1 &  -9.4 & -10.6 & -11.9 & -13.1 & -14.4 & -15.6 & -16.9 & -18.1 \\
$Y_{\rm MB}$&  -2.2 &   -4.1 &   -5.3 &   -3.9 &   -4.6 &   -4.3 &   -4.4 &   -3.2 &   -2.5 &   -2.1 &   -1.2 \\
$\sigma_Y$& 1.7 &  2.1 &  1.4 &  2.3 &  1.1 &  1.7 &  1.7 &  2.1 &  2.2 &  1.1 &  1.6 \\
\hline
\end{tabular}
\label{tab:rrl}
\end{table*}

%% file: lmc_bridge_table_yms.tex
\begin{table*}
\caption{Properties of the YMS bridge.}
\begin{tabular}{|c c c c c c c c|}
\hline
$X_{\rm MB}$& -7.5 &  -9.3 & -11.1 & -12.9 & -14.7 & -16.5 & -18.3 \\
$Y_{\rm MB}$&   0.0 &    1.3 &    0.2 &    0.9 &    0.3 &    0.1 &   -0.2 \\
$\sigma_Y$& 0.5 &  1.1 &  0.4 &  1.0 &  0.9 &  0.9 &  1.0 \\
\hline
\end{tabular}
\label{tab:yms}
\end{table*}